\documentclass[showpacs,aps,twocolumn,nofootinbib,11]{revtex4}
\usepackage{amsmath}
\usepackage{epsfig}
\usepackage{subfigure}
\usepackage{float}
\usepackage{longtable}
\usepackage{times}
\usepackage{txfonts}
\usepackage{color}
\usepackage{ulem}

 \usepackage{ifpdf}
 \ifpdf
 \usepackage[pdftex]{hyperref}
 \else
 \usepackage[hypertex]{hyperref}
 \fi

 \hypersetup{
   pdftitle={},%
   pdfauthor={},%
   pdfsubject={},%
   pdfkeywords={},%
   pdfstartview={},%
   bookmarksopen=true, breaklinks=true, debug=true, %
   colorlinks=true, linkcolor=red, citecolor=blue, urlcolor=blue
 }

\newcommand{\beq}{\begin{eqnarray}}
\newcommand{\eeq}{\end{eqnarray}}
\newcommand{\be}{\begin{eqnarray*}}
\newcommand{\ee}{\end{eqnarray*}}

\newcommand{\ie}{{\it i.e.}}

\newcommand{\ce}[1]{Eq.~\eqref{#1}}

\newcommand{\cf}[1]{{Fig.~\ref{#1}}}
\newcommand{\ct}[1]{{Table~\ref{#1}}}

\def\lsim{\raise0.3ex\hbox{$<$\kern-0.75em\raise-1.1ex\hbox{$\sim$}}}
\def\gsim{\raise0.3ex\hbox{$>$\kern-0.75em\raise-1.1ex\hbox{$\sim$}}}

\def\CuCu {CuCu}

\def\dAu  {$d$Au}

\def\AuAu {AuAu}
\def\pp   {$pp$}
\def\pA   {$pA$}
\def\AA   {$AA$}

\def\sqrtsNN {\mbox{$\sqrt{s_{NN}}$}}

\def\Ncoll   {\mbox{$N_{\rm coll}$}}

\def\RdAu    {\mbox{$R_{d\rm Au}$}}

\def\jpsi    {\mbox{$J/\psi$}}
\def\pT      {\mbox{$P_{T}$}}
\def\beq     {\begin{equation}}
\def\eeq     {\end{equation}}

\long\def\symbolfootnote[#1]#2{\begingroup%
  \def\thefootnote{\fnsymbol{footnote}}\footnote[#1]{#2}\endgroup}

\begin{document}


\title{Centrality, Rapidity and Transverse-Momentum Dependence of Cold Nuclear Matter Effects on \jpsi\ Production in 
\dAu\ , \CuCu\ and \AuAu\ Collisions at $\sqrt{s_{NN}}=200$ GeV}

\author
{E. G. Ferreiro$^a$, F. Fleuret$^b$, J.P. Lansberg$^{c,d}$\protect\footnote{Present address at Ecole polytechnique.}, A. Rakotozafindrabe$^e$
}

\affiliation{
$^a$ Departamento de F{\'\i}sica de Part{\'\i}culas, Universidade de Santiago de Compostela, 15782 Santiago de Compostela, Spain\\
$^b$Laboratoire Leprince Ringuet, \'Ecole polytechnique, CNRS/IN2P3,  91128 Palaiseau, France\\
$^c$Centre de Physique Th\'eorique, \'Ecole polytechnique, CNRS,  91128 Palaiseau, France\\
$^d$SLAC National Accelerator Laboratory, Theoretical Physics, Stanford University, Menlo Park, CA 94025, USA\\
$^e$IRFU/SPhN, CEA Saclay, 91191 Gif-sur-Yvette Cedex, France
}

\begin{abstract}
We have carried out a wide study of Cold Nuclear Matter (CNM) effects on \jpsi\ production in 
\dAu\ , \CuCu\ and \AuAu\ collisions at $\sqrt{s_{NN}}=200$ GeV. We have studied 
the effects of three different gluon-shadowing parametrisations, using the usual simplified kinematics
for which the momentum of the gluon recoiling against the \jpsi\ is neglected as well as an exact kinematics
for a $2\to 2$ process, namely $g+g\to J/\psi+g$ as expected from LO pQCD.
We have shown that the rapidity distribution of the nuclear modification factor $R_{dAu}$, 
and particularly its anti-shadowing peak, is systematically shifted toward larger rapidities in 
the  $2\to 2$ kinematics, irrespective of which shadowing parametrisation is used. In turn, we 
have noted differences in the effective final-state nuclear 
absorption needed to fit the PHENIX \dAu\ data. Taking advantage of our implementation of a
$2\to 2$ kinematics, we have also computed the transverse momentum
dependence of the nuclear modification factor, which cannot be predicted 
with the usual simplified kinematics. All the corresponding observables have been computed for 
\CuCu\ and \AuAu\ collisions and compared to the PHENIX and STAR data. Finally, 
we have extracted the effective nuclear 
absorption from the recent measurements of $R_{CP}$ in \dAu\ collisions by the PHENIX collaboration.
\end{abstract}

\maketitle



\section{Introduction}
\label{sec:intro}

The \jpsi\ particle is considered to be one of the most interesting probes 
of the transition
from hadronic matter to a deconfined state of QCD matter,
the so-called Quark-Gluon Plasma (QGP).
In the presence of a QGP, binding of $c{\bar c}$ ̄ pairs into  \jpsi\ mesons
is predicted to be hindered due to color screening,
leading to a \jpsi\ suppression in heavy ion collisions~\cite{Matsui86}.

The results of the \jpsi\ production in \AuAu\ 
collisions at \sqrtsNN=200~GeV~\cite{Adare:2006ns}
show a significant suppression.
Nevertheless, PHENIX data on \dAu\ collisions \cite{Adare:2007gn} have also
shown a non trivial behaviour,
pointing out that Cold Nuclear Matter (CNM) effects play an essential
role at these energies (for recent reviews see~\cite{Frawley:2008kk,Rapp:2008tf,Kluberg:2009wc}
and for perspectives for the LHC see~\cite{Lansberg:2008zm}).

All this reveals that
the interpretation of the results obtained in nucleus-nucleus collisions relies on a good
understanding and a proper subtraction of the CNM  effects, known to
impact the \jpsi\ production in deuteron-nucleus collisions, where the
deconfinement cannot be reached.

In previous studies~\cite{OurIntrinsicPaper,OurExtrinsicPaper},  
we have shown  that the impact of the gluon shadowing
on \jpsi\ production does depend on the partonic
process producing the $c \bar c$ and then the \jpsi.
Indeed, the evaluation of the shadowing corrections in which one treats the production as a
2$\rightarrow$2 process ($g+g\rightarrow \jpsi+g$) shows visible differences 
in the nuclear modifications factors when compared to the results obtained for a 2$\rightarrow$1 process.

The former  partonic production mechanism seems to be favoured by the recent 
studies based on Colour-Singlet Model (CSM), including~\cite{Haberzettl:2007kj} 
or not~\cite{Brodsky:2009cf} $s$-channel cut contributions.
This also seems to be confirmed by the recent studies of higher-order QCD corrections, which have 
shown, on one hand, that the problematic $P_T$ dependence of the LO 
CSM~\cite{CSM_hadron} is cured when going at $\alpha_S^4$ and 
$\alpha_S^5$~\cite{Campbell:2007ws,Artoisenet:2007xi,Gong:2008sn,Artoisenet:2008fc,Lansberg:2008gk} and, 
on the other,  that the CSM yield at NLO in $e^+e^-\to J/\psi+X_{{\rm non}\ c \bar c}$~\cite{Gong:2009kp,Ma:2008gq} 
saturates the experimental values by the Belle collaboration~\cite{Pakhlov:2009nj}. 
This does not allow for a significant Colour-Octet (CO) component~\cite{Zhang:2009ym}, which happens to
be precisely the one appearing in the low-$P_T$ description of hadroproduction via a $2\to 1$ process~\cite{Cho:1995ce}.
To summarise, one is entitled to consider that the former $2\to 2$ kinematics is the most appropriate to account 
for the PHENIX data.

The structure of this paper is as follows. In section II, we describe our approach, namely the partonic
process, the shadowing parametrisations and the implementation of the nuclear absorption that we have chosen. 
In section III, we present and discuss our results for $R_{dAu}$ versus rapidity, centrality and transverse
momentum. We particularly discuss the impact of the partonic process kinematics. Section IV is devoted to the 
results in \AuAu\ and \CuCu\ collisions. In section V, 
we present and discuss our extraction of the effective $J/\psi$ absorption cross section from the PHENIX \dAu\ data and we
finally conclude.


\section{Our approach}
\label{sec:ourapproach}


To describe the \jpsi\ production in nucleus collisions, our Monte-Carlo
framework~\cite{OurIntrinsicPaper,OurExtrinsicPaper} is based on the probabilistic Glauber model, the nuclear density
 profiles being defined with the Woods-Saxon parametrisation for
any nucleus ${A>2}$ and the Hulthen wavefunction for the
deuteron~\cite{Hodgson:1971}. The nucleon-nucleon inelastic cross section at
$\sqrtsNN=200\mathrm{~GeV}$ is taken to $\sigma_{NN}=42\mathrm{~mb}$ and the
maximum nucleon density to $\rho_0=0.17\mathrm{~nucleons/fm}^3$.

\subsection{Partonic process for $c \bar c$ production}

Most 
studies of CNM effects on \jpsi\ production~\cite{OtherShadowingRefs} 
rely on the assumption that the $c \bar c$ 
pair is produced by a $2\to 1$ partonic process where both initial particles are  gluons
 carrying some intrinsic transverse momentum~$k_T$. The sum of the gluon intrinsic transverse momentum 
is transferred to the $c \bar c$ pair, thus to the \jpsi\ since the soft hadronisation process 
does not significantly modify the kinematics. This is supported by the picture of the 
Colour Evaporation Model (CEM) at LO (see~\cite{Lansberg:2006dh} and references 
therein) or of the CO mechanism  at 
$\alpha^2_S$~\cite{Cho:1995ce}.
In such approaches, the transverse momentum $P_T$ of the \jpsi\ is meant to come {\it entirely}  from the intrinsic
transverse momentum of the initial gluons.

As just discussed, recent CSM-based studies~\cite{Haberzettl:2007kj,Brodsky:2009cf} have shown agreement with the 
PHENIX \pp\ data~\cite{Adare:2006kf} and the problematic $P_T$ dependence of the LO CSM has been shown 
to be cured at Tevatron energies when going at $\alpha_S^4$ and 
$\alpha_S^5$~\cite{Campbell:2007ws,Artoisenet:2007xi,Gong:2008sn,Artoisenet:2008fc,Lansberg:2008gk}.
Furthermore, $e^+e^-\to J/\psi+X_{{\rm non}\ c \bar c}$ NLO computations in the CSM~\cite{Gong:2009kp,Ma:2008gq} leave now
too small a room for a CO component which would support a $2\to 1$ production kinematics~\cite{Cho:1995ce}.

Parallel to this, intrinsic transverse momentum is not sufficient to describe the $P_T$ spectrum of quarkonia produced in
hadron collisions~\cite{Lansberg:2006dh}. For $P_T$ above approximately 2-3 GeV, one expects that 
most of the transverse momentum should have an extrinsic
origin, \ie\ the \jpsi's $P_T$ would be balanced by the emission of a recoiling particle in the final
state. The \jpsi\ would then be produced by gluon fusion in a $2\to 2$ process with emission of a hard final-state gluon. 
This emission has a definite influence on the kinematics of the
\jpsi\ production. Indeed, for a given \jpsi\ momentum (thus for
fixed~$y$ and $P_T$),  the process $g+g \to J/\psi +g$ will proceed on average from gluons with larger Bjorken-$x$
than  $g+g \to c\bar c \to J/\psi \,(+X)$. Therefore,
 they will be affected by different shadowing corrections. From now on, we will refer to the latter $2\to 1$
scenario as the {\it intrinsic} scheme, and to the former $2\to 2$ as the {\it extrinsic} scheme.

In the intrinsic scheme, we use the fits to the rapidity $y$ and \pT\ spectra measured
by PHENIX~\cite{Adare:2006kf} in \pp\ collisions at $\sqrt{s_{NN}}=200\mathrm{~GeV}$
as inputs of the Monte-Carlo.
The
measurement of the \jpsi\ momentum completely fixes the longitudinal
 momentum fraction carried by the initial partons:
\begin{equation}
x_{1,2} = \frac{m_T}{\sqrt{s_{NN}}} \exp{(\pm y)} \equiv x_{1,2}^0(y,P_T),
\label{eq:intr-x1-x2-expr}
\end{equation}
with the transverse mass $m_T=\sqrt{M^2+P_T^2}$, $M$ being the \jpsi\ mass.

On the other hand, in the extrinsic scheme, information from the data alone
-- the $y$ and \pT\ spectra -- is not sufficient to determine $x_1$ and $x_2$.
Indeed, the presence of a final-state particle introduces further degrees of freedom in the kinematics
allowing more than one value of $(x_1, x_2)$ for a given set $(y, P_T)$ -- which are ultimately the measured quantities as
opposed to $(x_1, x_2)$ .
 The quadri-momentum conservation explicitly results in a more complex expression of $x_2$ as a function of~$(x_1,y,P_T)$:
\begin{equation}
x_2 = \frac{ x_1 m_T \sqrt{s_{NN}} e^{-y}-M^2 }
{ \sqrt{s_{NN}} ( \sqrt{s_{NN}}x_1 - m_T e^{y})} \ .
\label{eq:x2-extrinsic}
\end{equation}
Equivalently, a similar expression can be written for $x_1$ as a function of~$(x_2,y,P_T)$.
In this case, models are mandatory to compute the proper
weighting of each kinematically allowed $(x_1, x_2)$. This weight is simply
the differential cross section at the partonic level times the gluon Parton Distribution 
Functions (PDFs),
\ie\ $g(x_1,\mu_f) g(x_2, \mu_f) \, d\sigma_{gg\to J/\psi + g} /dy \, dP_T\, dx_1 dx_2 $.
In the present implementation of our code, we are able to use the partonic differential
cross section computed from {\it any} theoretical approach. For now, we use the one
from~\cite{Haberzettl:2007kj} which takes into account the $s$-channel cut
contributions~\cite{Lansberg:2005pc} to the basic CSM and
satisfactorily describes the data down to very low~\pT~\ .
A study using other matrix elements (LO CSM, NLO CEM,\dots) is planned for future works.

\begin{figure}[htb!]
\includegraphics[width=1.0\linewidth]{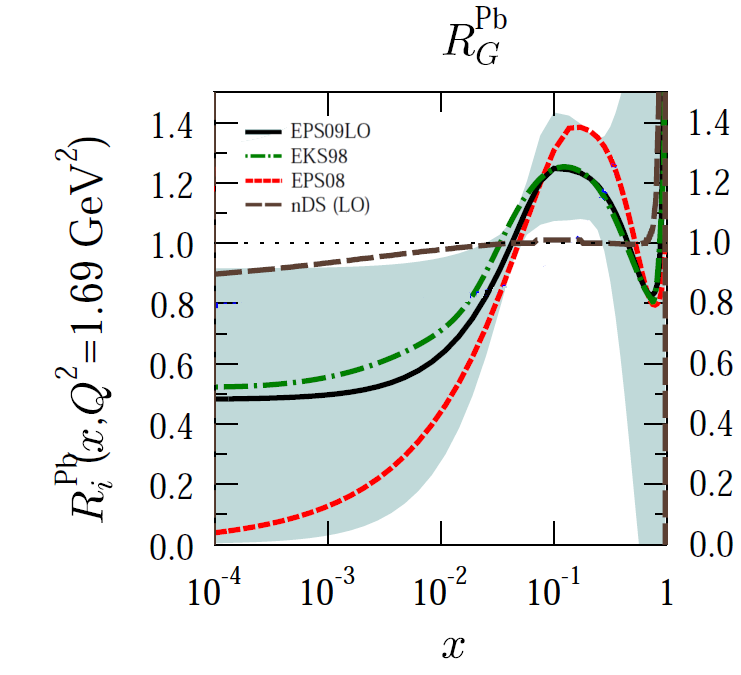}
\caption{(Color online) comparison of the gluon shadowing parametrisations EKS98~\cite{Eskola:1998df}, EPS08~\cite{Eskola:2008ca},
nDS~\cite{deFlorian:2003qf} at LO and EPS09~\cite{Eskola:2009uj} at LO in a Lead nucleus at $Q^2=1.69$ GeV$^2$. Adapted from~\cite{Eskola:2009uj}.}
\label{fig:EPS09vsothers}
\end{figure}

\begin{figure*}[htb!]
\begin{center}
\includegraphics[width=1.0\linewidth]{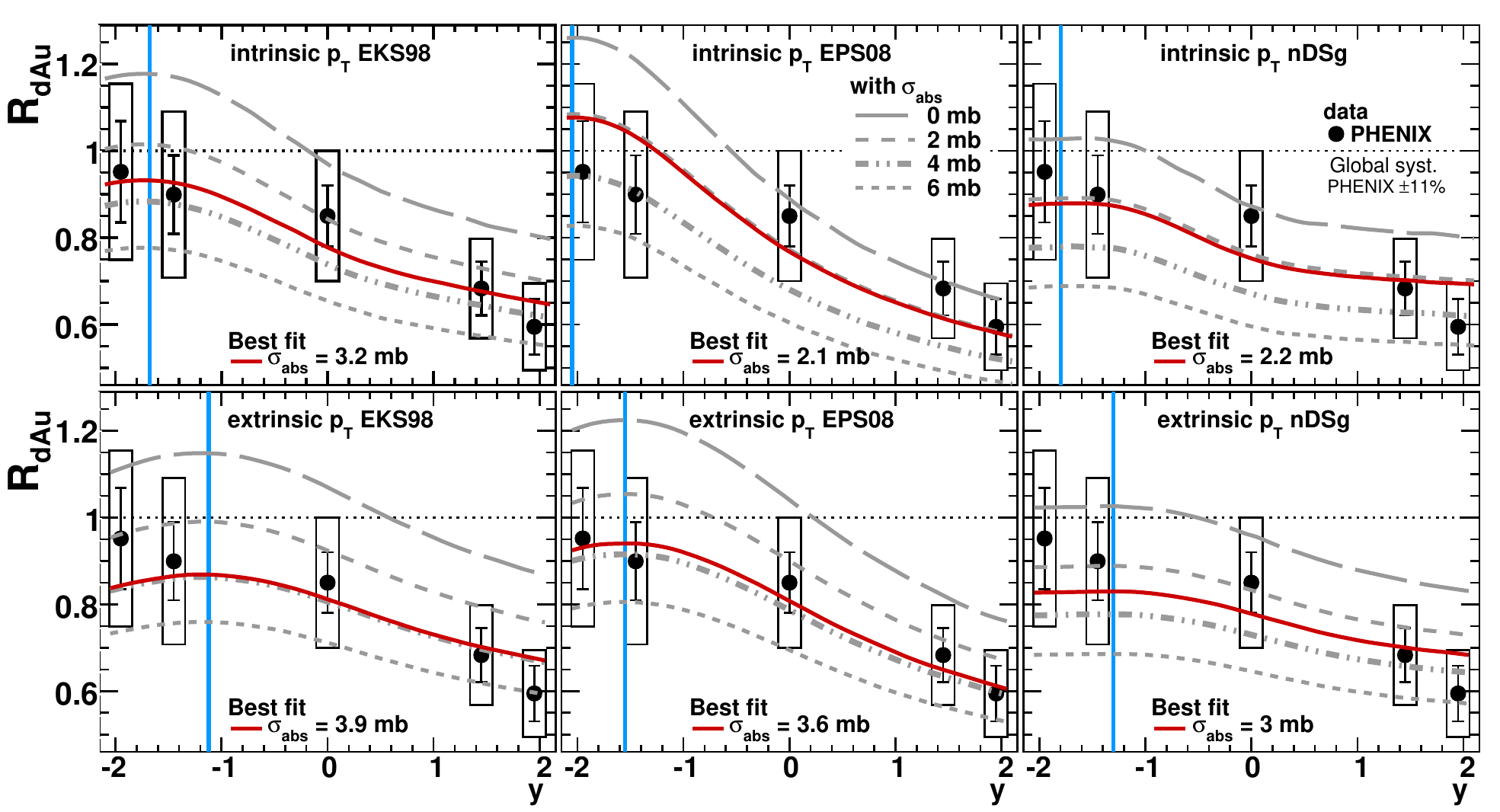}
\end{center}
\caption{(Color online) \jpsi\ nuclear modification factor in \dAu\ collisions, $R_{dAu}$, at $\sqrt{s_{NN}}=200\mathrm{~GeV}$ versus $y$ in the 
intrinsic (top row) and extrinsic (lower row) schemes using
the following gluon shadowing parametrisations: EKS98~\cite{Eskola:1998df} (left), EPS08~\cite{Eskola:2008ca} (middle), nDSg~\cite{deFlorian:2003qf} (right). The vertical lines indicate qualitatively the antishadowing peak and its shifts towards larger $y$ in the extrinsic scheme.
}
\label{fig:RdAu_vs_y}
\end{figure*}

\subsection{Shadowing}

To get the \jpsi\ yield in \pA\ and \AA\ collisions, a shadowing-correction
factor has to be applied to the \jpsi\ yield obtained from the simple
superposition of the equivalent number of \pp\ collisions.
This shadowing factor can be expressed in terms of the ratios $R_i^A$ of the
nuclear Parton Distribution Functions (nPDF) in a nucleon of a nucleus~$A$ to the
PDF in the free nucleon. 

These parametrisations provide the nuclear ratios $R_i^A$
of the PDFs: 
\beq
\label{eq4}
R^A_i (x,Q^2) = \frac{f^A_i (x,Q^2)}{ A f^{nucleon}_i (x,Q^2)}\ , \ \
i = q, \bar{q}, g \ .
\eeq
The numerical parametrisation of $R_i^A(x,Q^2)$
is given for all parton flavours. Here, we restrain our study to gluons since, at
high energy, \jpsi\ is essentially produced through gluon fusion \cite{Lansberg:2006dh}.

We shall consider three
different shadowing parametrisations for comparison: EKS98~\cite{Eskola:1998df}, EPS08~\cite{Eskola:2008ca} and
nDSg~\cite{deFlorian:2003qf} at LO.
Recently, a new parametrisation, EPS09~\cite{Eskola:2009uj}, has been made available. It offers the possibility of 
properly taking into account the errors arising from the fit procedure. Yet, in the case of gluon nPDF,
as illustrated by~\cf{fig:EPS09vsothers}, the region spanned by this parametrisation is approximately bounded 
by both the nDS and EPS08 ones. However, we shall not consider the nDS parametrisation 
which shows such small shadowing corrections that no significant yield corrections are expected. 
We shall prefer to use nDSg
as done in other works since it provides a lower bound of EPS09 in the antishadowing region. 
Furthermore, the central curve of EPS09 is very close to EKS98.
In this context, we have found it reasonable to limit our studies to EKS98, EPS08 and nDSg and to postpone to a further
study the analysis of the error correlations and their impacts of $J/\psi$ production studies\footnote{Note that, as shown in~\cf{fig:EPS09vsothers}, the EPS08 parametrisation shows a strong gluon shadowing at very small $x$ due to the inclusion of forward-rapidity BRAHMS data in the fit. An explanation for such an effect has been based on the idea that, in this kinematic region that corresponds to the beam fragmentation region at large Feynman $x_F$, one can reach the smallest values of the momentum fraction variable $x_2$ in nuclei. It makes it possible to access the strongest coherence effects such as those associated with shadowing or alternatively the Color Glass Condensate (CGC). Nevertheless, as it has been argued in \cite{Kopeliovich:2008}, although at forward rapidities one accesses the smallest $x_2$ in the nuclear target, one simultaneously gets into the region of large $x_1$ of the projectile nucleon where energy conservation becomes an issue. Indeed, as stated in \cite{Kopeliovich:2005}, at large $x_F$ (or $x_1$) one expects a suppression from Sudakov form factors, giving the probability that no particle be produced as $x_F \to 1$, as demanded by energy conservation. In a pA collision, the multiple interactions of the nucleon remnants with the nucleus makes this less likely to occur and the suppression is expected to be stronger. In fact, factorisaton breaks down and the effective parton distributions in the projectile nucleon then become correlated with the nucleus target, see Eq. (15) of \cite{Kopeliovich:2005}. This effect should not be confused with gluon shadowing or other manifestations of coherence. Because of this, the strength of EPS08 gluon shadowing may be overestimated, since it includes data at large $x_F$ which were analysed assuming that the suppression was attributed only to a reduction of the gluon distribution in the nucleus. However, we would like to emphasize that, while the gluon shadowing in EPS08 and EKS98 differ by a factor 3 at $Q^2=1.69$ GeV$^2$, the difference is already less than 20$\%$ at $Q^2 \simeq M_{J/\psi}^2$ which is the relevant scale for our analysis.}.\\
Note that the spatial dependence of the shadowing has been included in our approach, assuming an inhomogeneous shadowing proportional to the local
density~\cite{Klein:2003dj,Vogt:2004dh}.

\begin{figure*}[thb!]
\subfigure[~EKS98]{\includegraphics[width=0.33\textwidth]{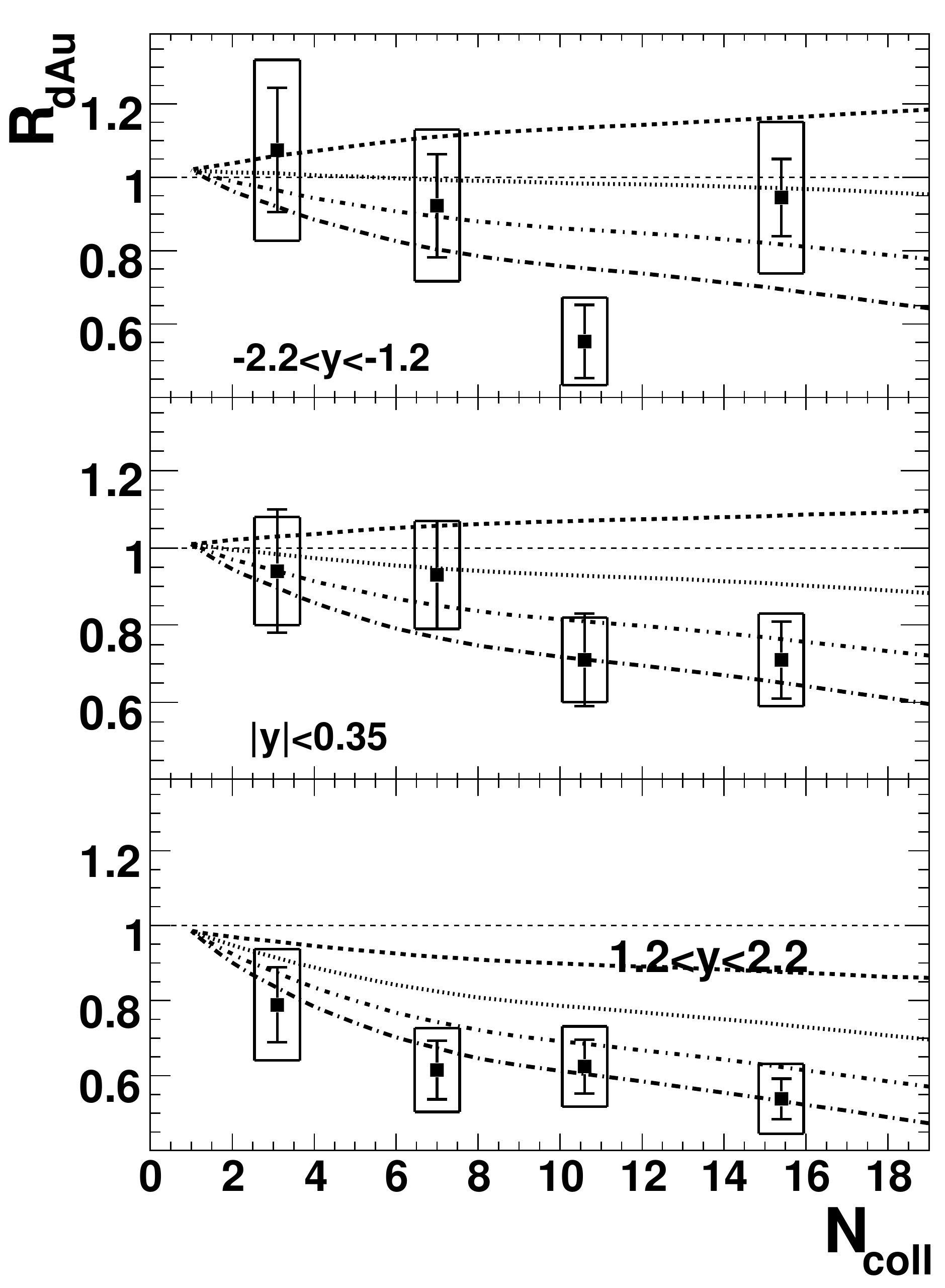}}
\subfigure[~EPS08]{\includegraphics[width=0.33\textwidth]{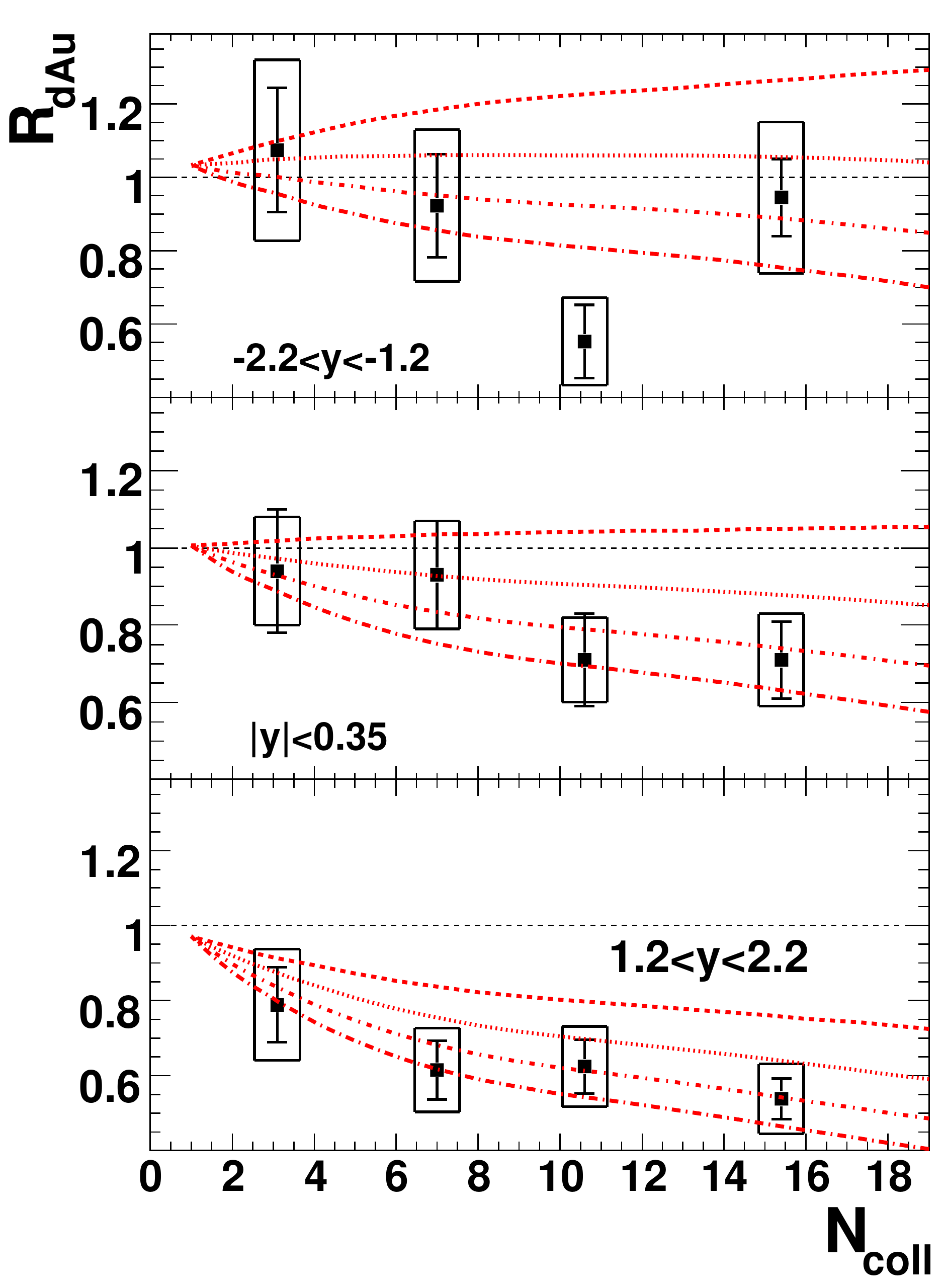}}
\subfigure[~nDSg]{\includegraphics[width=0.33\textwidth]{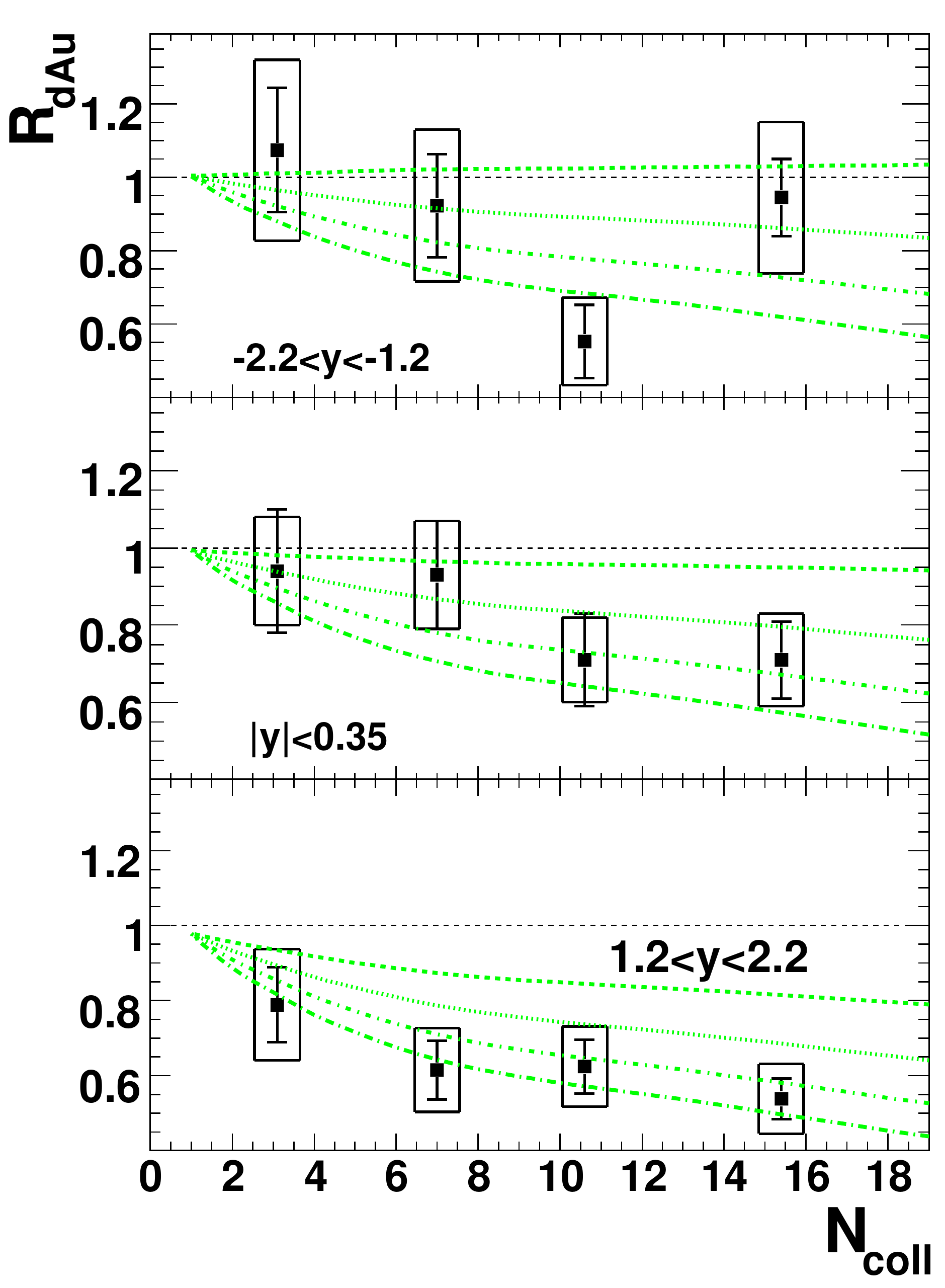}}
\caption{(Color online) \jpsi\ nuclear modification factor in \dAu\ collisions, $R_{dAu}$, at $\sqrt{s_{NN}}=200\mathrm{~GeV}$ versus the number of collisions for three rapidity ranges and for four values of the nuclear absorption (from top to bottom: $\sigma_{abs}=$0, 2, 4 and 6 mb) using  a) EKS98, b)  EPS08 and c) nDSg in the extrinsic scheme.}
\label{fig:rdau_ncoll}
\end{figure*}

\subsection{The nuclear absorption}

The second CNM effect that we have taken into account concerns
the nuclear absorption.  In the framework of the probabilistic Glauber
model, this effect refers to the probability for the pre-resonant $c{\bar c}$
pair to survive the propagation through the nuclear medium and is usually parametrised
by an effective absorption cross section~$\sigma_{\mathrm{abs}}$.
Our results will be first shown for different values of $\sigma_{\mathrm{abs}}$
using the three aforementioned shadowing parametrisations. Afterwards, we will
extract the values that provides the best fit to the PHENIX data. We note here that 
this effective cross section may also account for initial state effects, such 
as parton energy loss in the nuclear target.

\section{Results for $\rm dAu$ collisions}

The \jpsi\ suppression is usually characterised by the {\it nuclear modification factor} $R_{AB}$, {\it i.e.},
the ratio obtained by normalising the \jpsi\ yield in ion collisions to the \jpsi\ yield in 
proton collisions at the same energy times the average
number of binary inelastic nucleon-nucleon collisions $N_{coll}$:
\beq
R_{AB}=\frac{dN_{AB}^{J/\psi}}{\langle\Ncoll\rangle dN_{pp}^{J/\psi}}.
\eeq
Any nuclear effect affecting $J/\psi$ production leads to a deviation of $R_{AB}$ from unity.

\begin{figure*}[thb!]
\subfigure[~EKS98]{\includegraphics[width=0.33\textwidth]{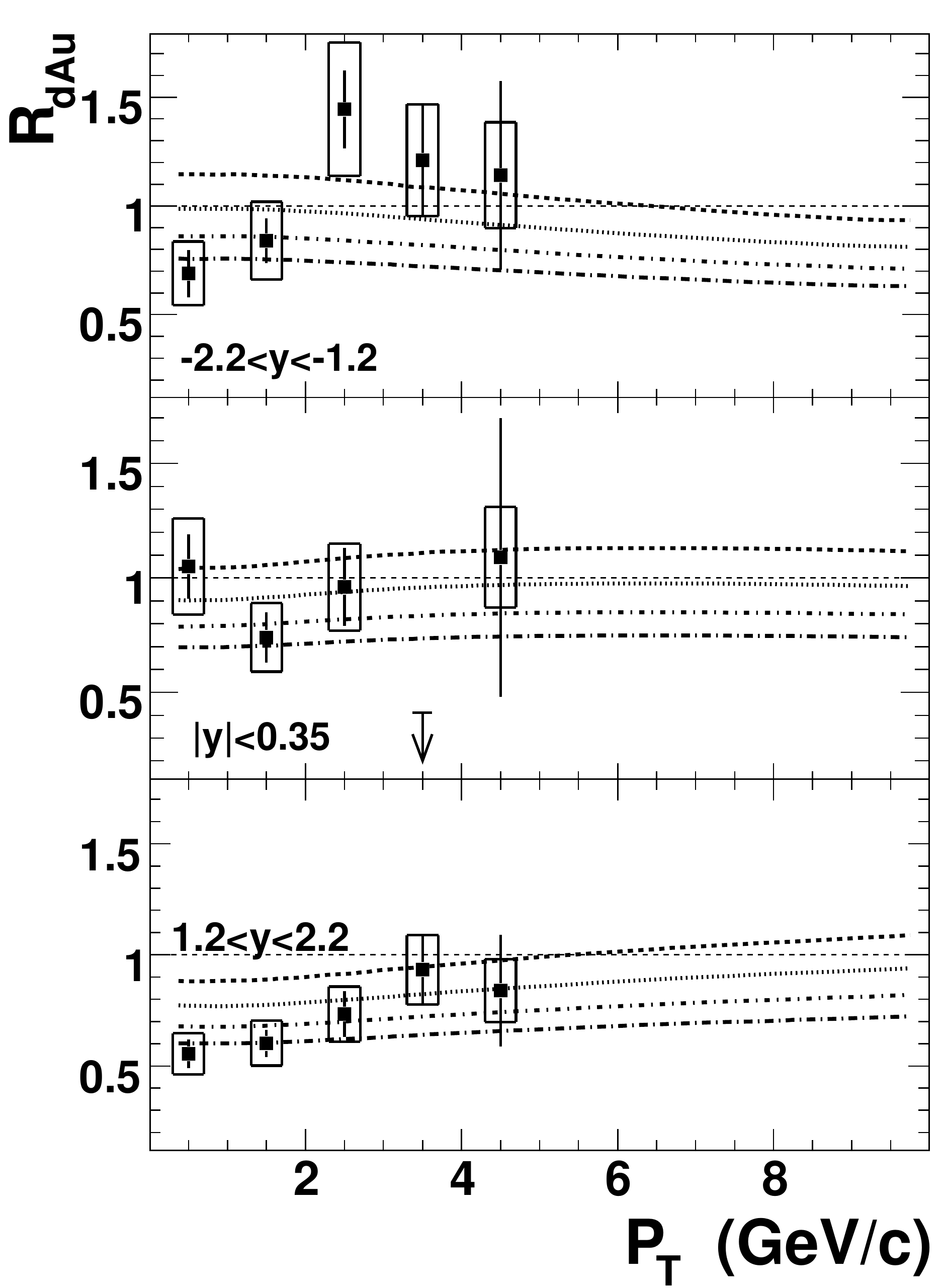}}
\subfigure[~EPS08]{\includegraphics[width=0.33\textwidth]{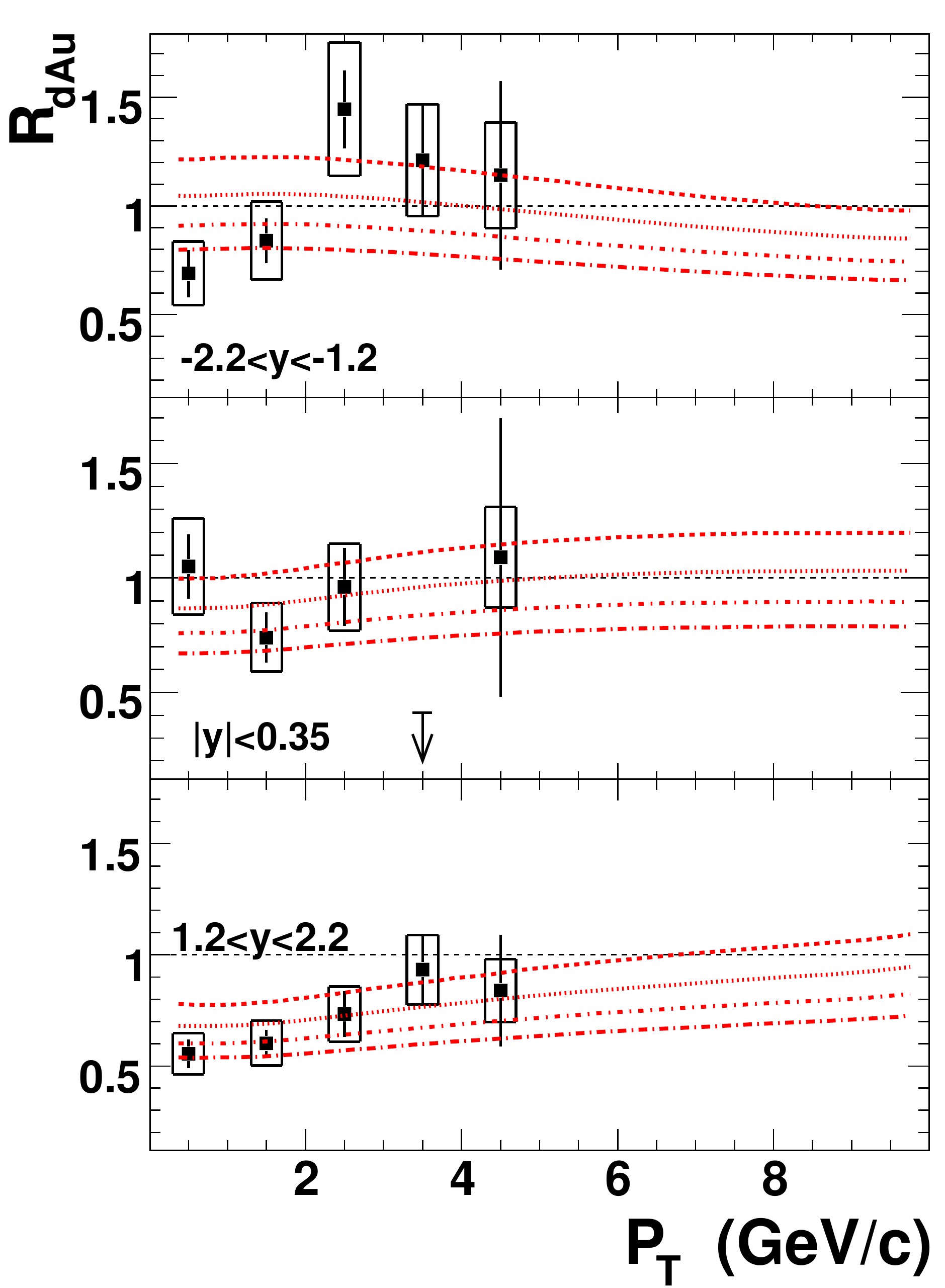}}
\subfigure[~nDSg]{\includegraphics[width=0.33\textwidth]{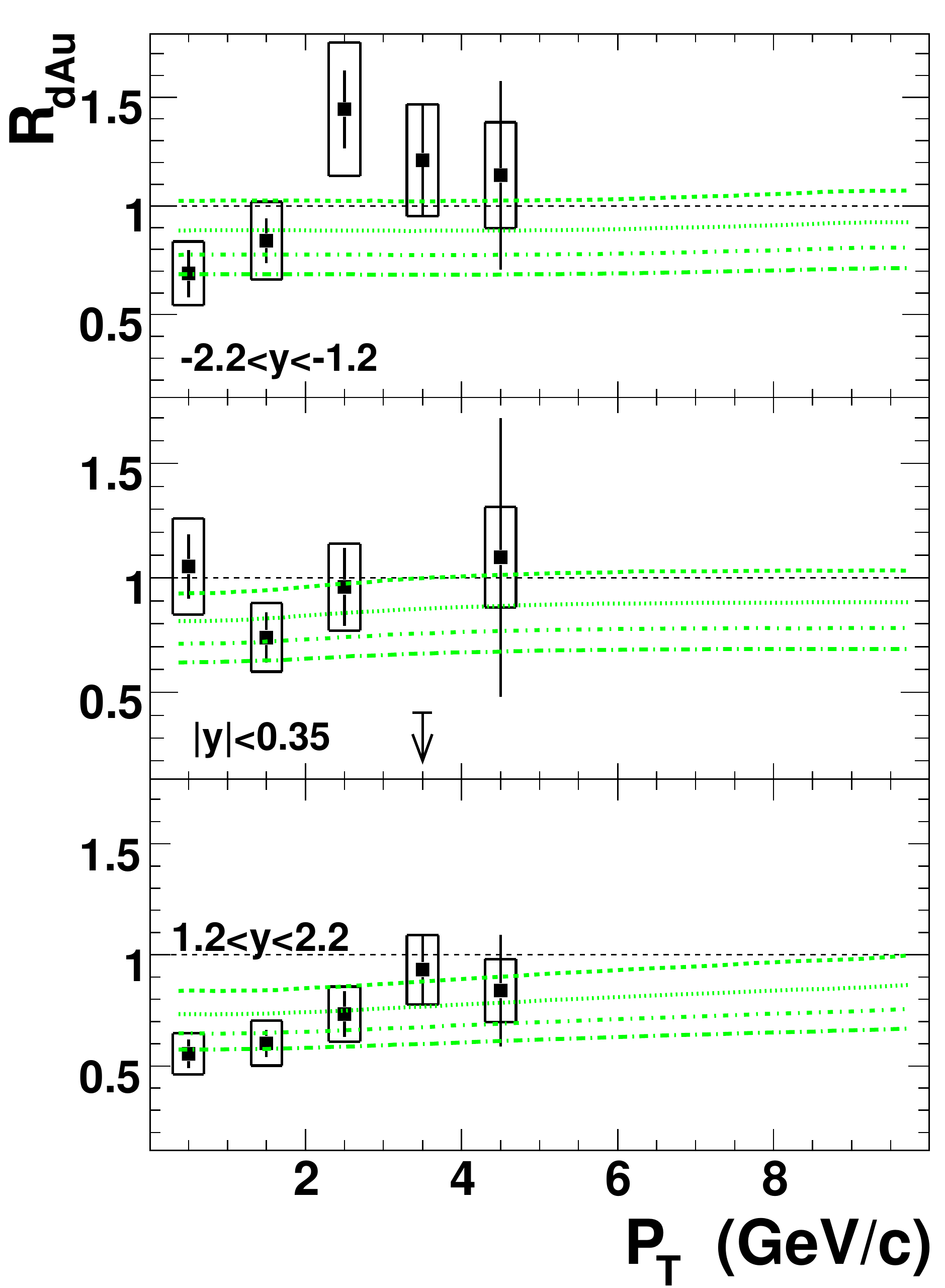}}
\caption{(Color online) \jpsi\ nuclear modification factor in \dAu\ collisions, $R_{dAu}$, at
$\sqrt{s_{NN}}=200\mathrm{~GeV}$ versus the transverse momentum for three rapidity ranges and for
four values of the nuclear absorption (from top to bottom: $\sigma_{abs}=$0, 2, 4 and 6 mb) using
 a) EKS98, b)  EPS08 and c) nDSg in the extrinsic scheme.}
\label{fig:rdau_pT}
\end{figure*}

\subsection{$R_{dAu}$ vs rapidity: distribution shift }

In Fig.~\ref{fig:RdAu_vs_y}, we show $R_{dAu}$ vs $y$ according to both the 
extrinsic and intrinsic schemes. The results are displayed 
for four values of $\sigma_{\mathrm{abs}}$ for each of the aforementioned shadowing parametrisations and
 are compared with the PHENIX data~\cite{Adare:2007gn}. The best fit result as performed in section \ref{sec:fit} 
 is also shown.

The comparison between the three plots on the upper row and their corresponding plots on the lower one
shows a striking -- but expectable -- feature : the rapidity distribution in the extrinsic scheme
is systematically shifted toward larger $y$ compared to the intrinsic case. This is 
particularly visible when one focuses on the anti-shadowing peak, which we have indicated 
qualitatively with vertical lines.

Such a shift is, in fact, not surprising at all. It simply reflects the larger value of 
the gluon momentum fraction in the nucleus, $x_2$, needed to produce a $J/\psi$ when the 
momentum of the final state gluon is explicitly taken into account in the computations.

As mentioned above, recent theoretical studies of $J/\psi$ production in vacuum, \ie~in $pp$ collisions, 
support at low and mid $P_T$ a partonic production mechanism as given by LO pQCD, 
namely $gg \to J/\psi g$, as opposed to a $2\to 1$ process.
In this context, we claim that this rapidity shift -- evident for any shadowing parametrisation --
is a feature of $J/\psi$ production in \dAu\ that should be systematically accounted for. 
Along the same lines of arguments, we shall focus in the following discussions on the results
obtained in the extrinsic scheme ($2\to 2$ case), except for the extraction of 
$\sigma_{abs}$ using the $R_{dAu}$ and $R_{CP}$ results.

\begin{figure*}[hbt!]
\subfigure[~EKS98]{\includegraphics[width=0.33\textwidth]{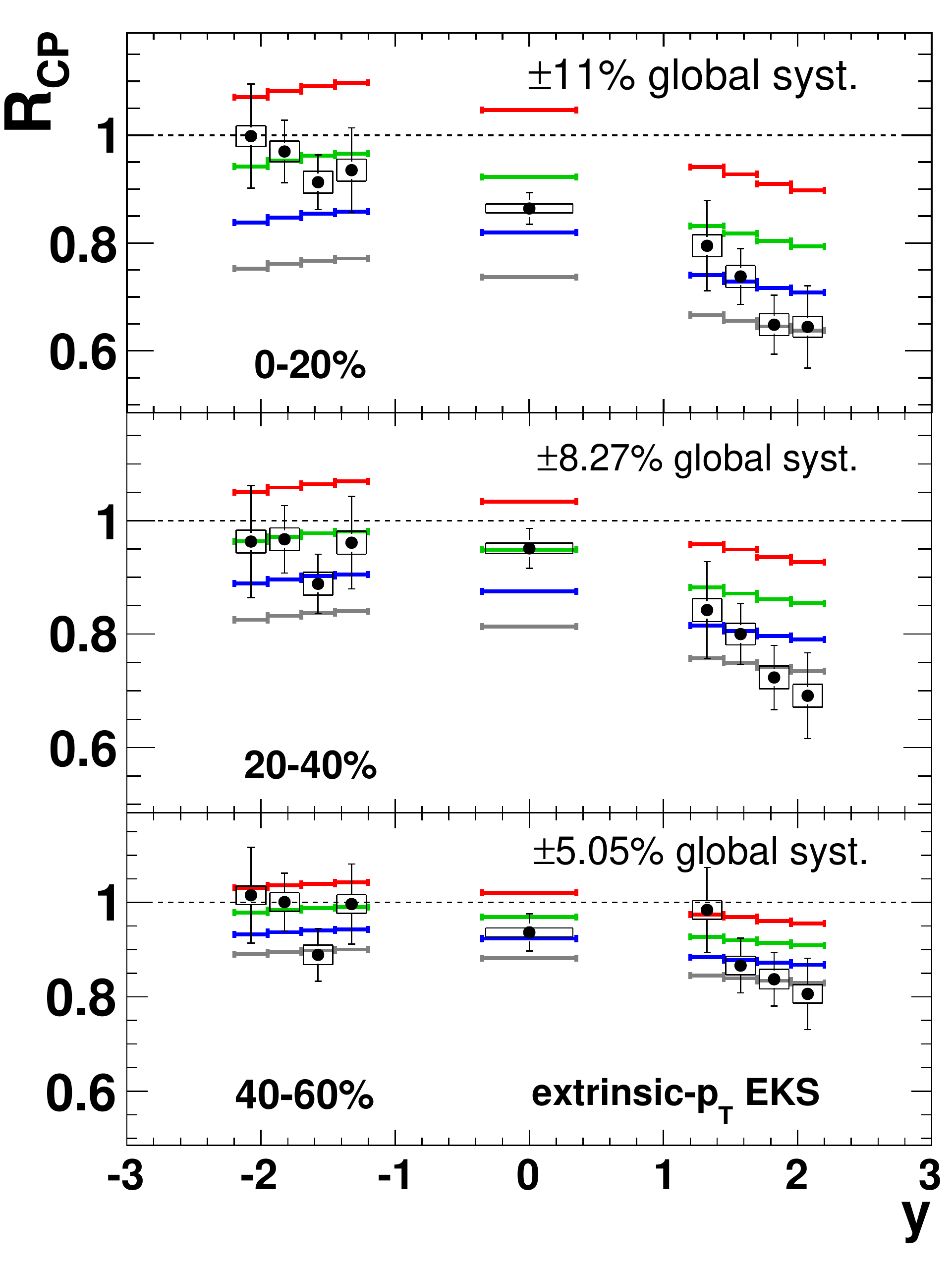}}
\subfigure[~EPS08]{\includegraphics[width=0.33\textwidth]{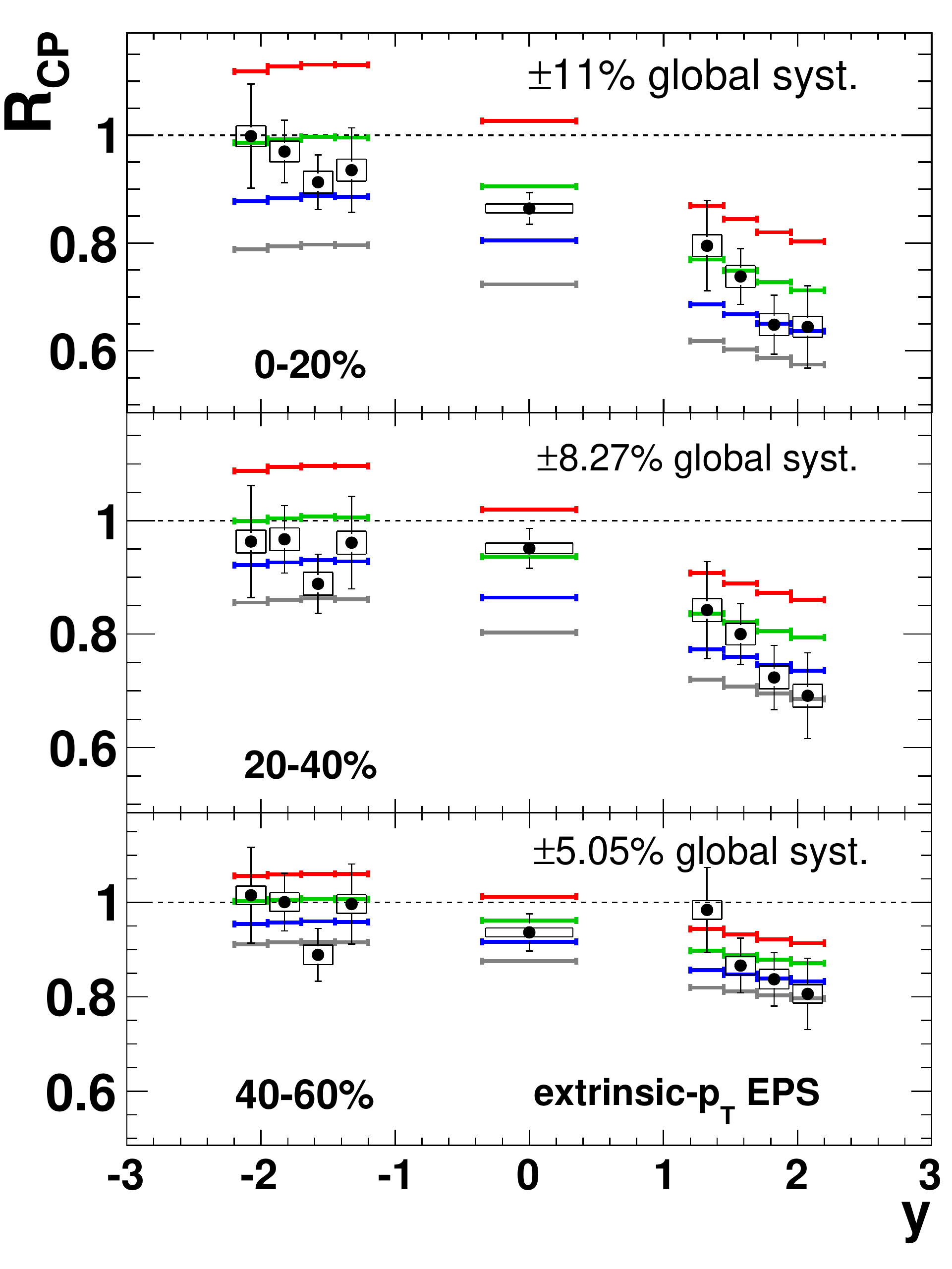}}
\subfigure[~nDSg]{\includegraphics[width=0.33\textwidth]{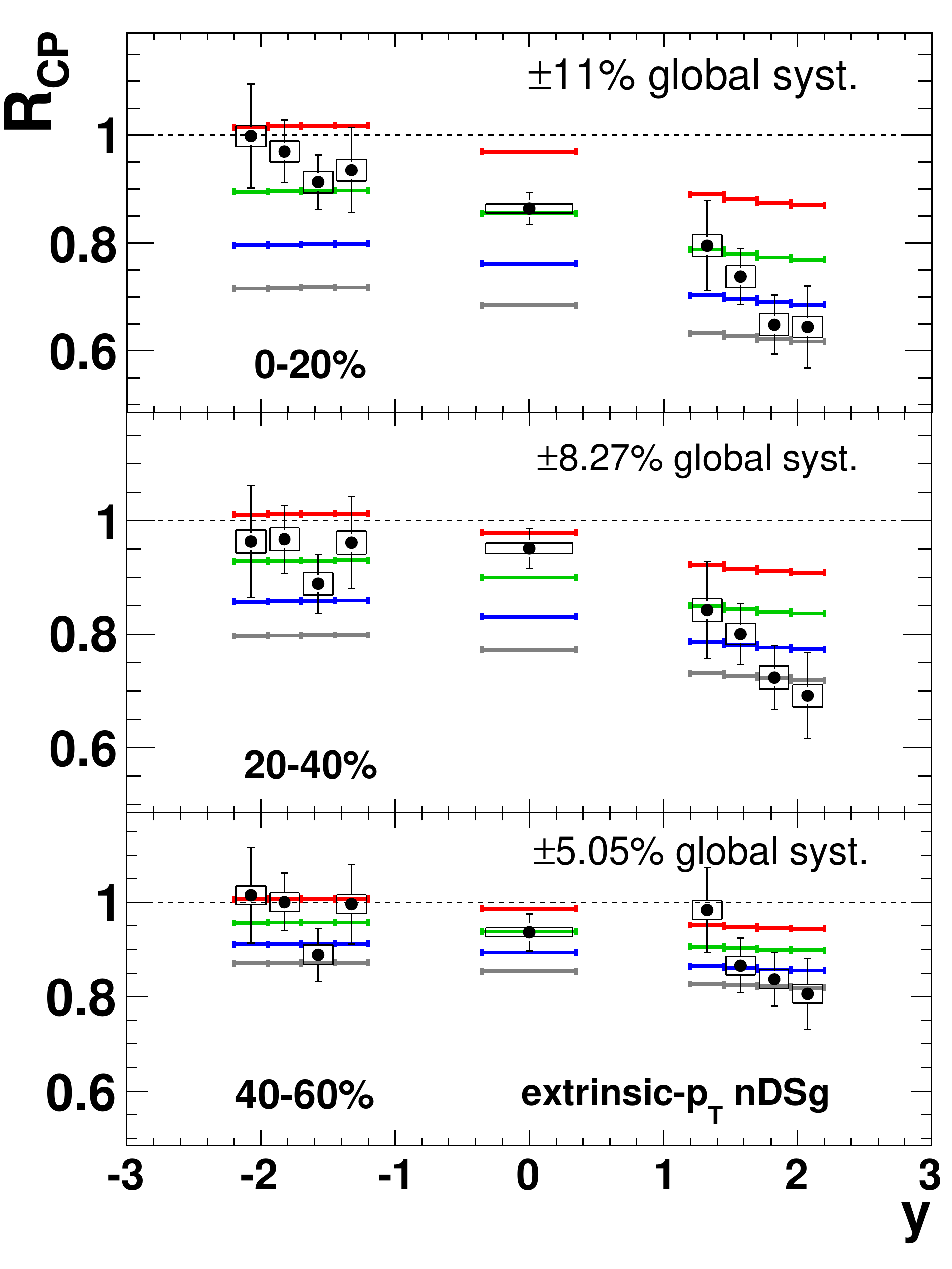}}
\caption{(Color online) \jpsi\ central to peripheral nuclear modification factor in \dAu\ collisions, $R_{CP}$, at $\sqrt{s_{NN}}=200\mathrm{~GeV}$ versus rapidity
for three centrality ranges and for four values of the nuclear absorption (from top to bottom: $\sigma_{abs}=$0, 2, 4 and 6 mb)
 using : a) EKS98, b)  EPS08, c) nDSg.}
\label{fig:rcp}
\end{figure*}

\subsection{$R_{dAu}$ vs centrality}

In Figs.~\ref{fig:rdau_ncoll} we present our results 
for $R_{dAu}$ versus centrality, expressed as the number of collisions. 
We have taken into account the three shadowing parametrisations and four values of $\sigma_{\mathrm{abs}}$.
One observes the effect of  the impact parameter dependence of the shadowing --increasing for
inner production points-- which induces a progressive increase  (resp. decrease) vs $N_{\rm coll}$ 
in the backward (resp. forward) due to the anti-shadowing (resp. shadowing) effect. Indeed,
for collisions with larger $N_{\rm coll}$, the \jpsi\ creation points are on average closer
to the gold nucleus center where the shadowing is expected to be stronger.
For the same reason, the absorption  suppresses the yield more strongly for larger $N_{\rm coll}$.

The overall effect (see Figs.~\ref{fig:rdau_ncoll}) matches the trend of the PHENIX data~\cite{Adare:2007gn}, 
showing a decrease vs $N_{\rm coll}$
stronger in the forward region than in the backward.

We also note that STAR collaboration has recently released a preliminary measurement of $R_{dAu}$ in the 
region $|y|<0.5$ using the most central collisions ($0-20 \%$)~\cite{Perkins:2009tp}: $R_{dAu}=1.45 \pm 0.60$. 
Higher statistics are however needed to draw 
conclusions from those data.

\subsection{$R_{dAu}$ vs  transverse momentum}

It is important to note that, in order to predict the transverse momentum dependence of the shadowing 
corrections, one needs to resort to a model which contains an explicit 
dependence on $P_T$. Studies were earlier carried on using the  CEM at NLO in~\cite{Bedjidian:2004gd}.
However, due to the complexity inherent to the NLO code used, it was not possible to implement the
impact parameter dependence of the shadowing, needed to reproduce, for instance,
the centrality dependence~\cite{Klein:2003dj,Vogt:2004dh} as just discussed.
Thanks to the versatility of our Glauber code, we can carry on such computation including
such an impact parameter dependence as well as involved production mechanisms containing 
a non-trivial dependence on $P_T$.

In Figs.~\ref{fig:rdau_pT},  we show our results on 
$R_{dAu}$ versus the transverse momentum. We emphasise that the growth
of $R_{dAu}$ is not related to any Cronin effect, since it is not taken into account here.
It simply comes from the increase of $x_2$ for increasing $P_T$ as given by \ce{eq:x2-extrinsic}. 
Hence, in the mid and forward rapidity region, $x_2$ goes through the anti-shadowing region and one 
observes an enhancement in $R_{dAu}$. In the backward region,
where $x_2$ sits in anti-shadowing region, 
one only sees a decrease. A similar behaviour is obviously expected in $R_{AA}$ vs $P_T$ 
as will be discussed in section \ref{subsec:RAAvsPT}.

~\\~\\~\\~\\~\\~\\

\subsection{$R_{CP}$}

    New results for the \jpsi\ from the 2008 \dAu\ run with approximately thirty times larger
   integrated luminosity than the 2003 \dAu\ results  are emerging,
   with the first preliminary result in terms of $R_{CP}$~\cite{daSilva:2009yy},
\begin{eqnarray}
  \label{eq:rcp}
  R_{CP} = \frac{\left(\frac{dN_{J/\psi}}{dy}/N_{coll}
    \right)}{\left(\frac{dN^{60-80\%}_{J/\psi}}{dy}/N^{60-80\%}_{coll}
    \right)}.
\end{eqnarray}

\begin{figure*}[hbt!]
\subfigure[~EKS98]{\includegraphics[width=0.33\textwidth]{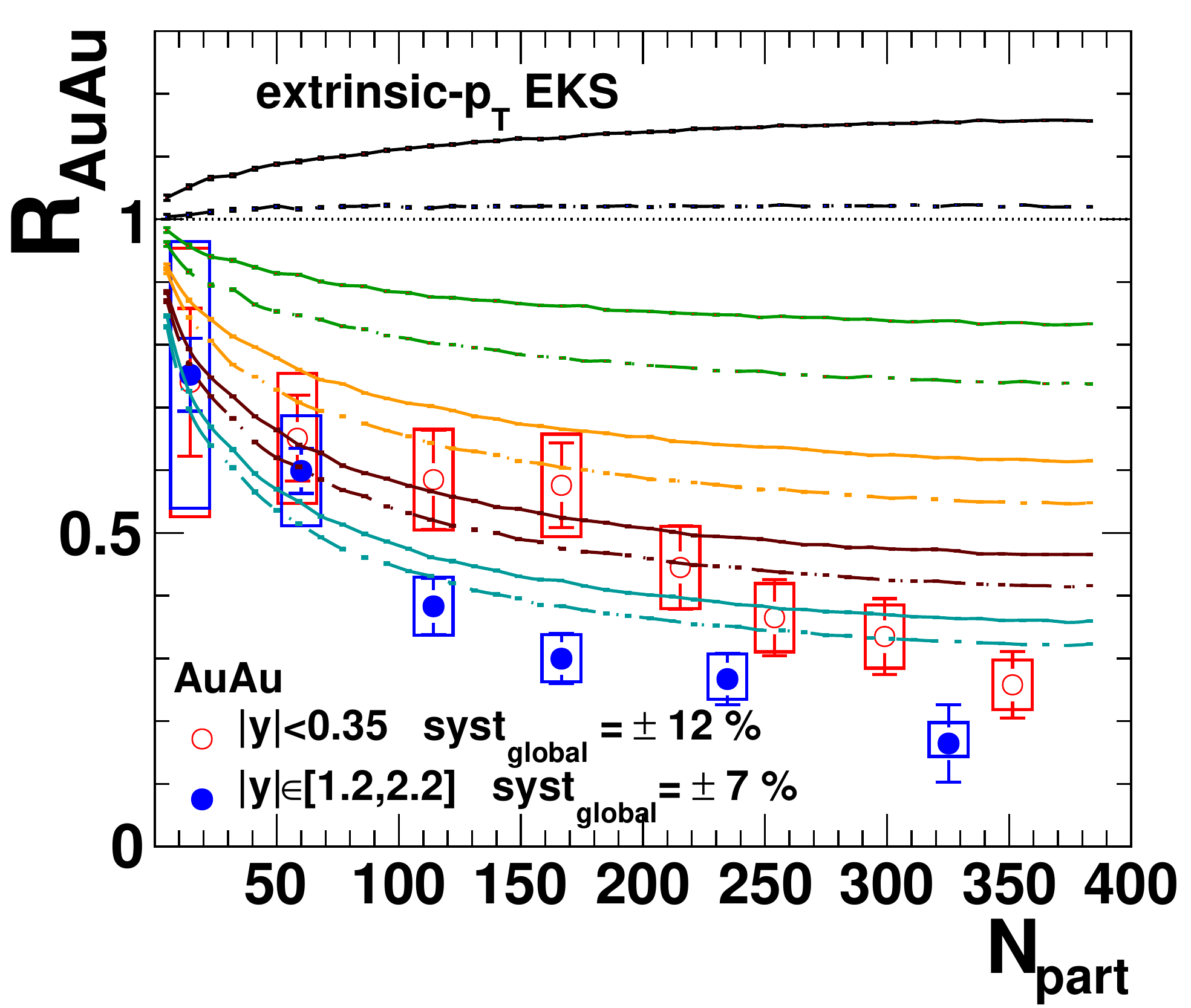}}
\subfigure[~EPS98]{\includegraphics[width=0.33\textwidth]{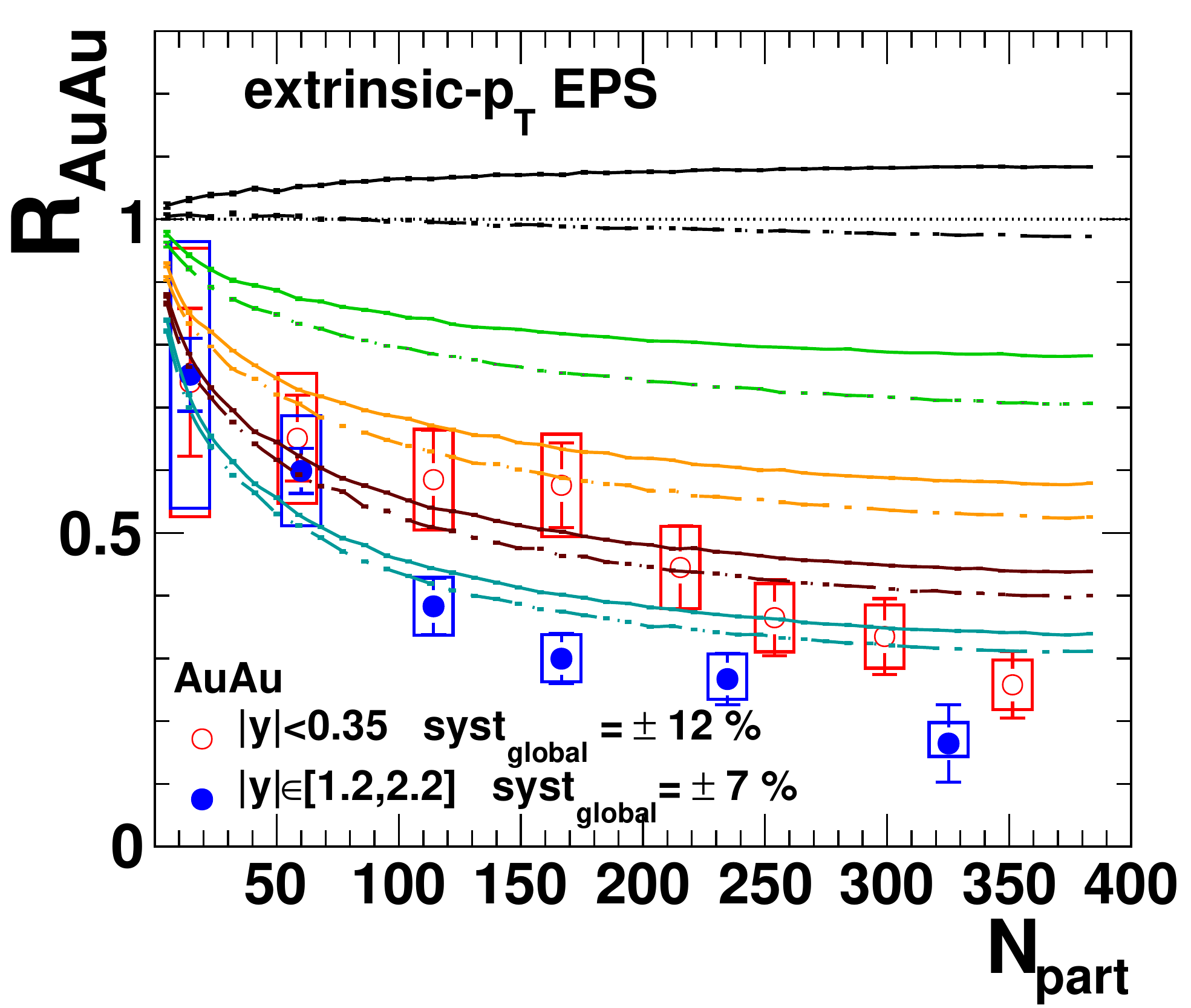}}
\subfigure[~nDSg]{\includegraphics[width=0.33\textwidth]{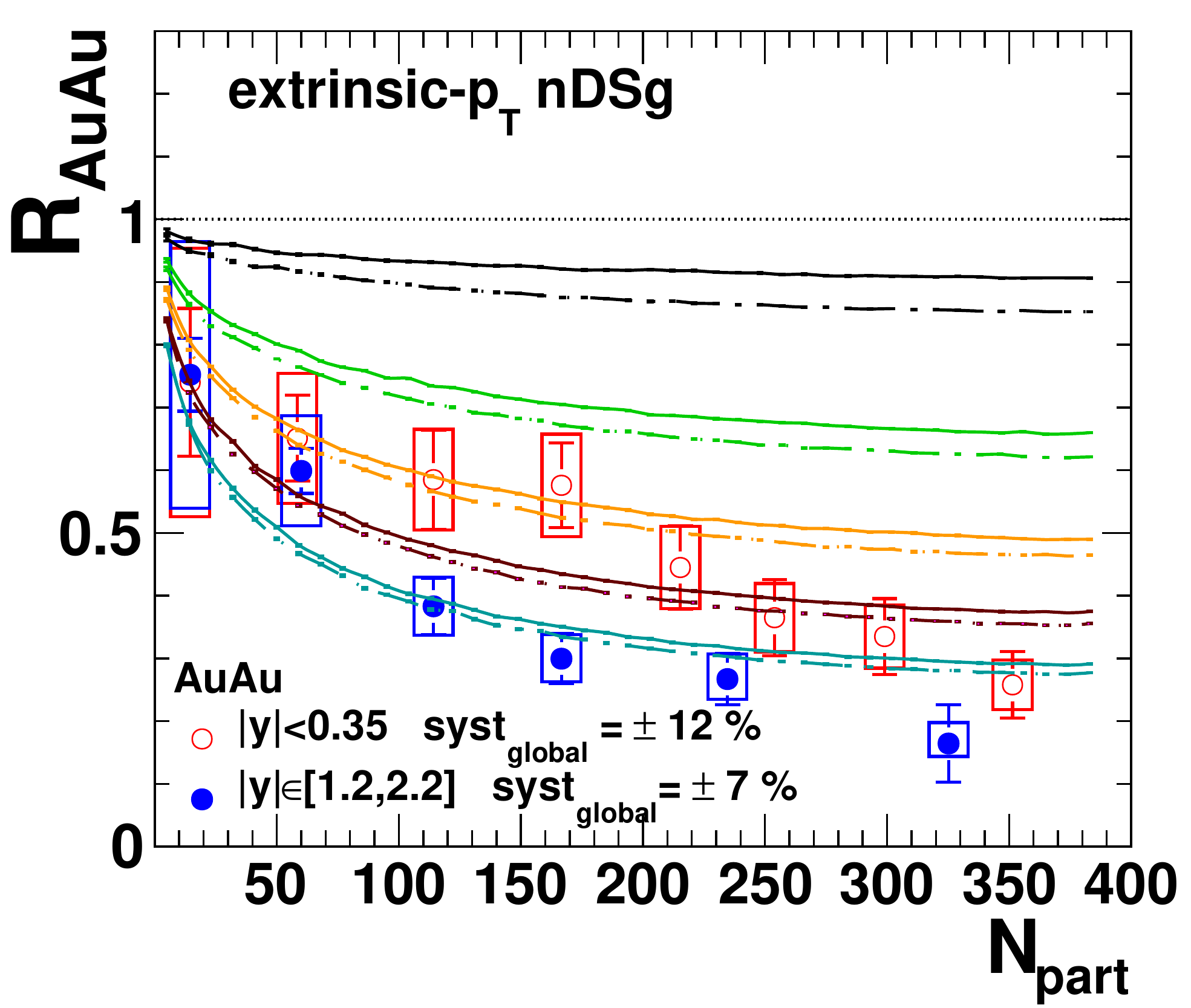}}
\caption{(Color online) centrality dependence of $R_{AuAu}^{\rm forward}$ (solid lines) and $R_{AuAu}^{\rm mid}$ (dashed lines) 
for five values of the nuclear absorption (from top to bottom: $\sigma_{\rm abs}=0$, 2, 4, 6 and 8 mb) using
 a) EKS98, b)  EPS08, c) nDSg, compared with the corresponding PHENIX data~\cite{Adare:2006ns}}
\label{fig:RAuAu_vs_Npart-cent_and_forw}
\subfigure[~EKS98]{\includegraphics[width=0.33\textwidth]{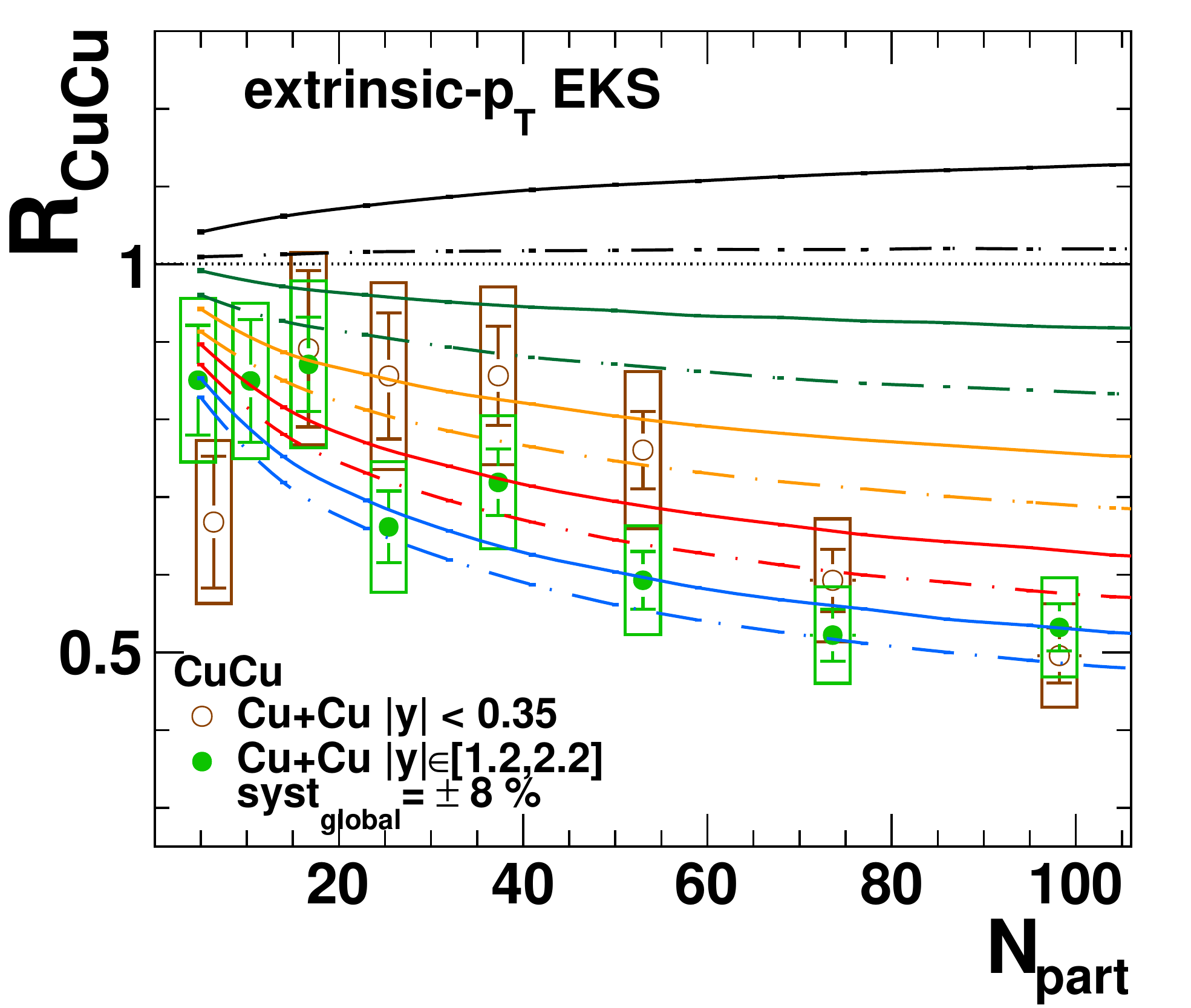}}
\subfigure[~EPS98]{\includegraphics[width=0.33\textwidth]{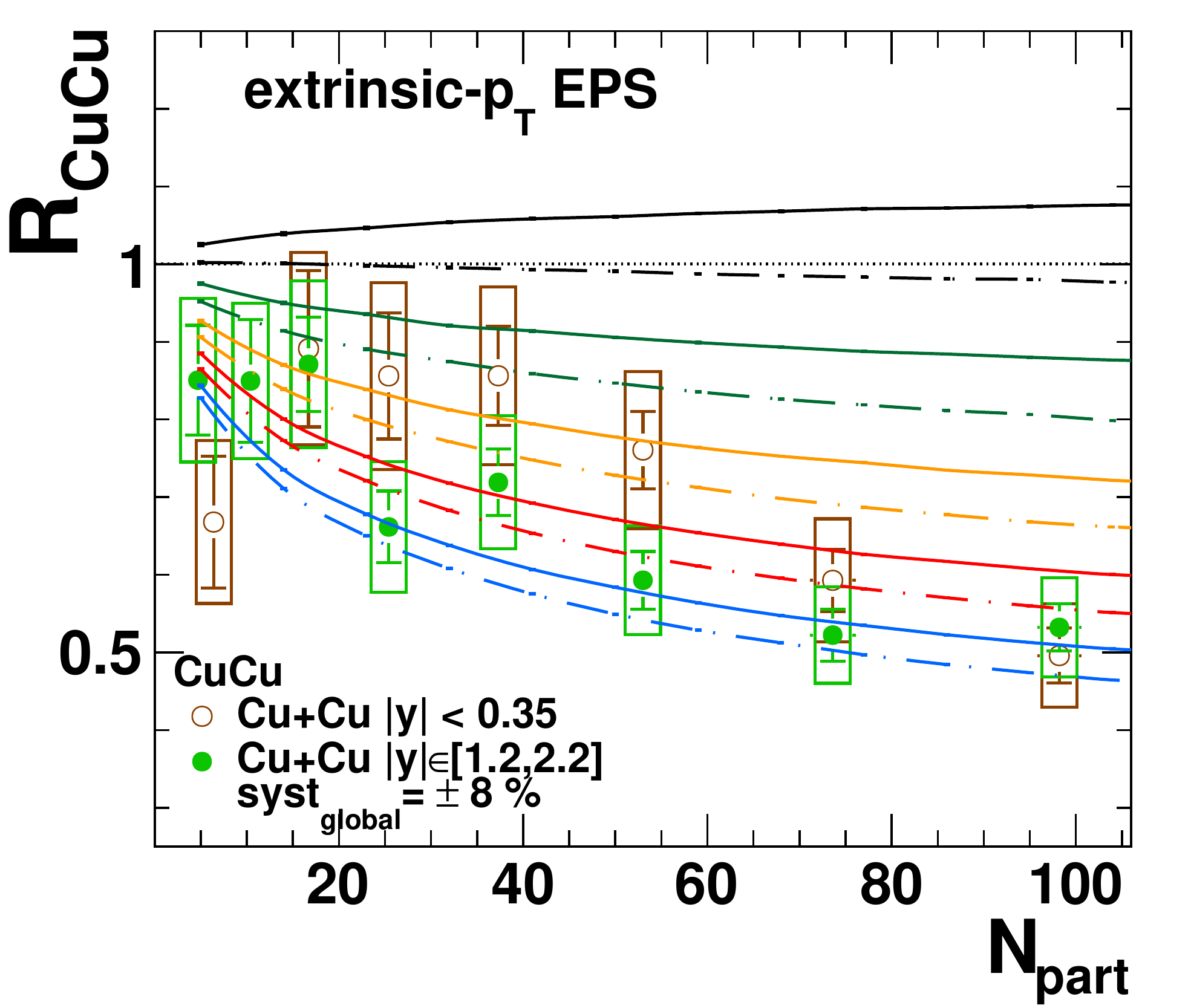}}
\subfigure[~nDSg]{\includegraphics[width=0.33\textwidth]{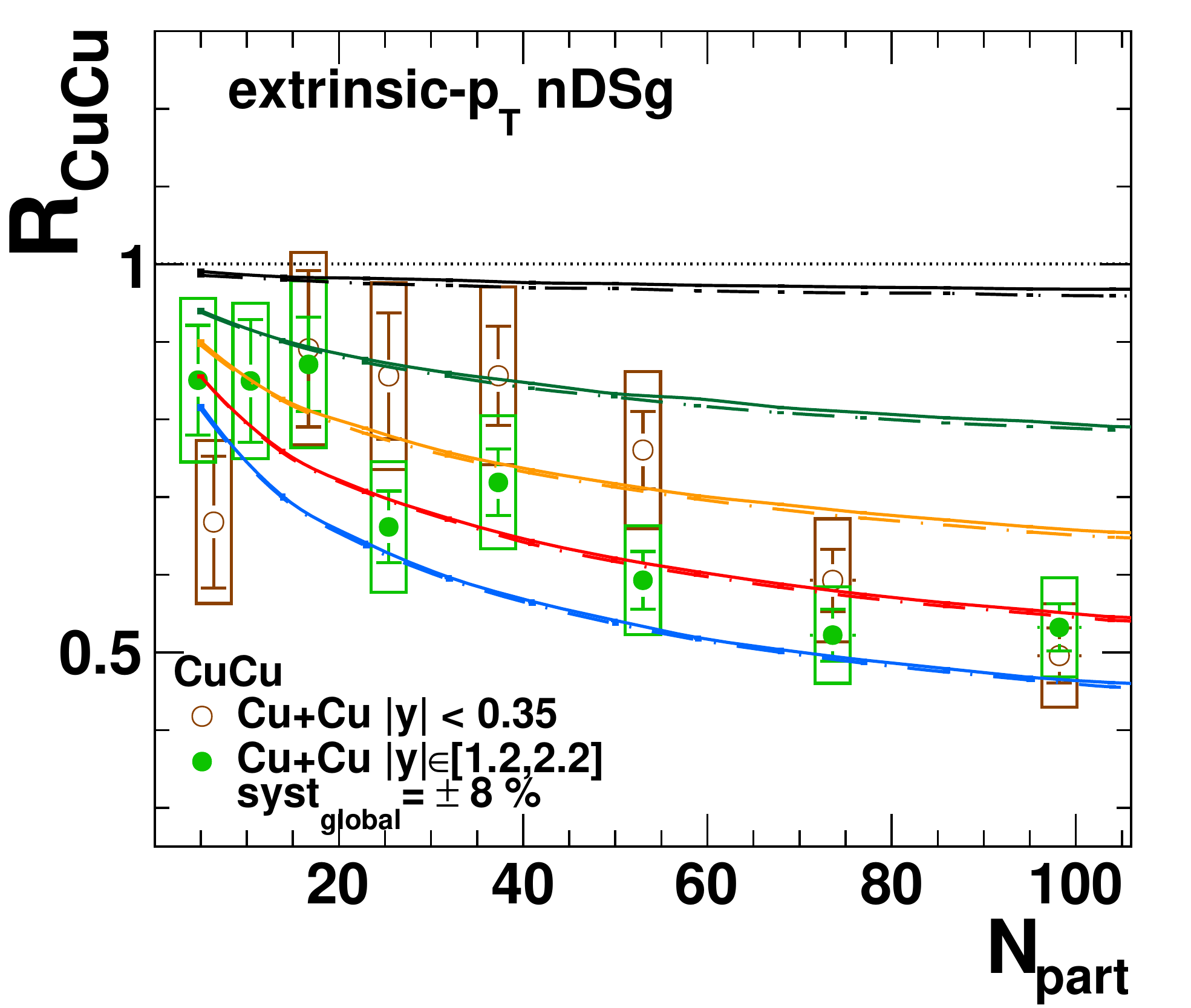}}
\caption{(Color online) centrality dependence of $R_{CuCu}^{\rm forward}$ (solid lines) and $R_{CuCu}^{\rm mid}$ (dashed lines) 
for five values of the nuclear absorption (from top to bottom: $\sigma_{abs}=$0, 2, 4, 6 and 8 mb) using
 a) EKS98, b)  EPS08, c) nDSg, compared with the corresponding PHENIX data~\cite{Adare:2006ns}.}
\label{fig:RCuCu_vs_Npart-cent_and_forw}
\end{figure*}

\begin{figure}[htb!]
\includegraphics[width=0.7\linewidth]{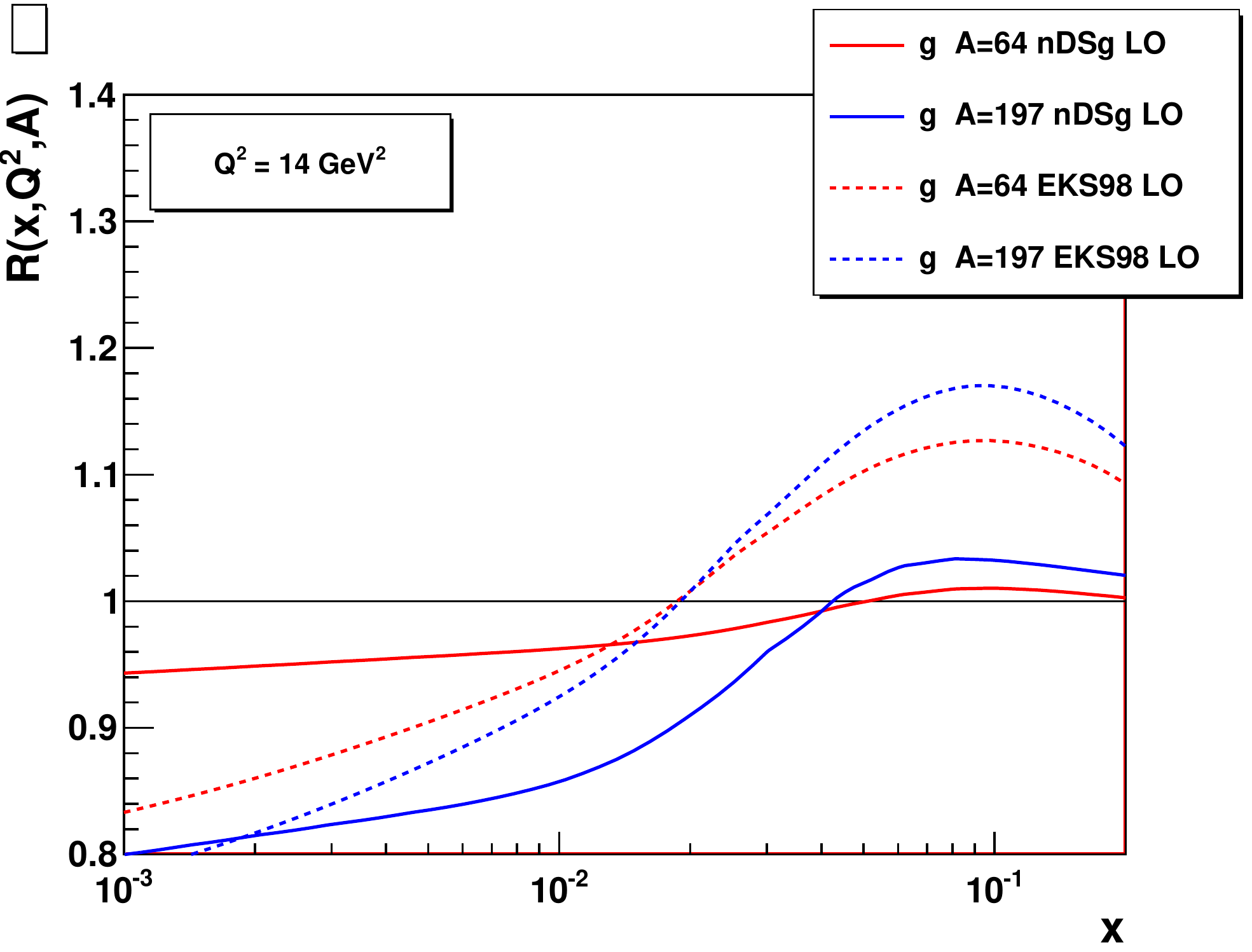}
\caption{(Color online) comparison of the gluon shadowing parametrisations EKS98~\cite{Eskola:1998df} and nDSg~\cite{deFlorian:2003qf}  
in lead and  copper nuclei at $Q^2=14$ GeV$^2$\protect\footnote{The plot has been generated by the nPDF generator
{\tt http://lappweb.in2p3.fr/lapth/npdfgenerator}}.}
\label{fig:shadowing-Cu}
\end{figure}

Those recent data show a significant dependence
on the rapidity, with a visible suppression for the most forward points in the three 
centrality ranges ($0-20\%$, $20-40\%$ and $40-60\%$). In the negative rapidity region, which
is expected to be dominated by large $x_2$ contributions\footnote{One recalls here that in the extrinsic
scheme there is no one-to-one mapping between the rapidity and the momentum fraction of the initial
gluon. Yet they are correlated.},
 the data (see \cf{fig:rcp}) show approximately no nuclear effects within the uncertainties, but a
visible suppression is observed at mid and forward rapidities. 
We also note that the  suppression is stronger in most central collisions. As regards our results, Fig. \ref{fig:rcp}
shows the rapidity dependence of the modification factor $R_{CP}$ 
for the three centrality ranges  using the EKS98, EPS08 and nDSg shadowing parametrisations. 
Different break-up cross sections have also been considered, from 0 to 6 mb.
As announced, we have chosen to focus on the extrinsic scheme. Like for $R_{dAu}$, this induces differences 
on the shadowing impact compared to an intrinsic scheme used in~\cite{frawley-INT}. One observes on~\cf{fig:rcp} that
the trend of the data is reasonably described with a $\sigma_{\rm abs}$ in the range of 2-4 mb, while 
the most forward points seem to decrease more than our evaluation. However, the uncertainties affecting the present preliminary
measurements  precludes drawing firmer conclusions. We shall analysis this in more detail
in section~\ref{sec:fit}.  

\section{Results for $\rm CuCu$ and $\rm AuAu$ collisions}

\subsection{Centrality dependence\protect\footnote{\rm As announced, we focus on the extrinsic case in all the following discussions.}}

In \cf{fig:RAuAu_vs_Npart-cent_and_forw} and \cf{fig:RCuCu_vs_Npart-cent_and_forw}, 
we present the centrality dependence 
of the nuclear
modification factor $R_{AuAu}$ and $R_{CuCu}$ in the forward and mid rapidity regions.
This has been computed for the three shadowing parametrisations and for five $\sigma_{\rm abs}$.

As we have already observed in~\cite{OurExtrinsicPaper}, $R_{AA}$ is systematically smaller
in the forward region that in the mid rapidity region (see also the next section). The difference increases
for more central collisions since we have used an impact parameter dependent shadowing.  While this difference 
(approximatively 20\% for rather central collisions) matches
well the one of the data when  $\sigma_{\mathrm{abs}}=0$,  one would have to choose a large
$\sigma_{\mathrm{abs}}$ if one wanted to reproduce the normalisation of the \AuAu\ data, disregarding any 
effects of Hot Nuclear Matter (HNM). However, for such large  $\sigma_{\mathrm{abs}}$, surviving $J/\psi$ 
from inner production points would be so rare that the difference between shadowing effects at mid
 and forward rapidities would nearly vanish. Note that for a $\sigma_{\mathrm{abs}}$ in the range of 2-4 mb, a difference remains.

While most of the above discussion is similar for CuCu collisions using the EKS and EPS shadowing, one observes a peculiar feature
for nDSg. Indeed, one does not observe any effect. This comes for the very weak shadowing encoded in nDSg for Cu. 
This is illustrated on~\cf{fig:shadowing-Cu}, where one sees that nDSg shadowing in Cu nucleus ends up to be very small.
Indeed, while there is a moderate difference between the EKS shadowing in Cu and in Au, 
there is a significant difference for nDSg. 

\begin{figure*}[thb!]
\subfigure[~EKS98]{\includegraphics[width=0.33\textwidth]{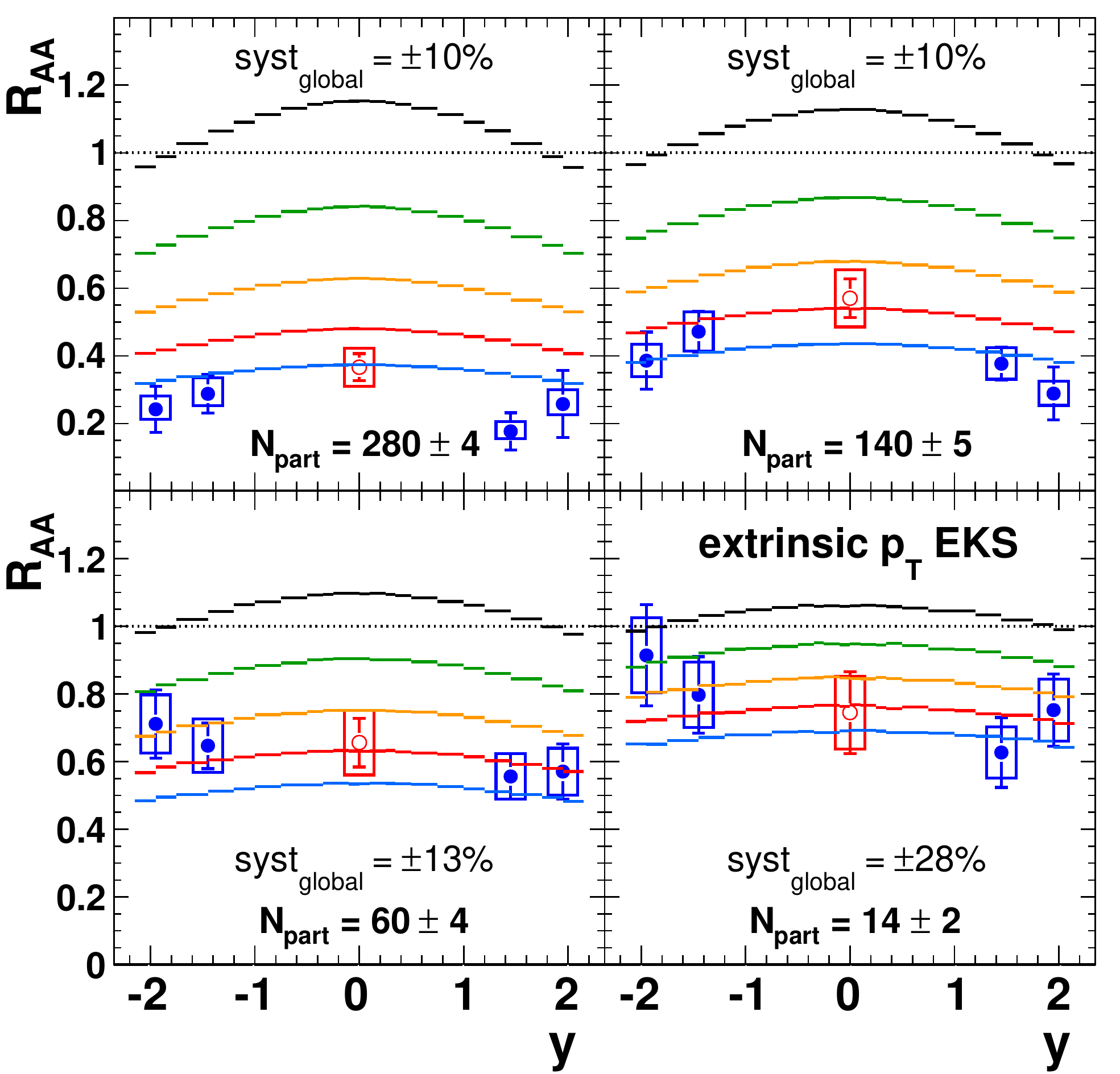}}
\subfigure[~EPS08]{\includegraphics[width=0.33\textwidth]{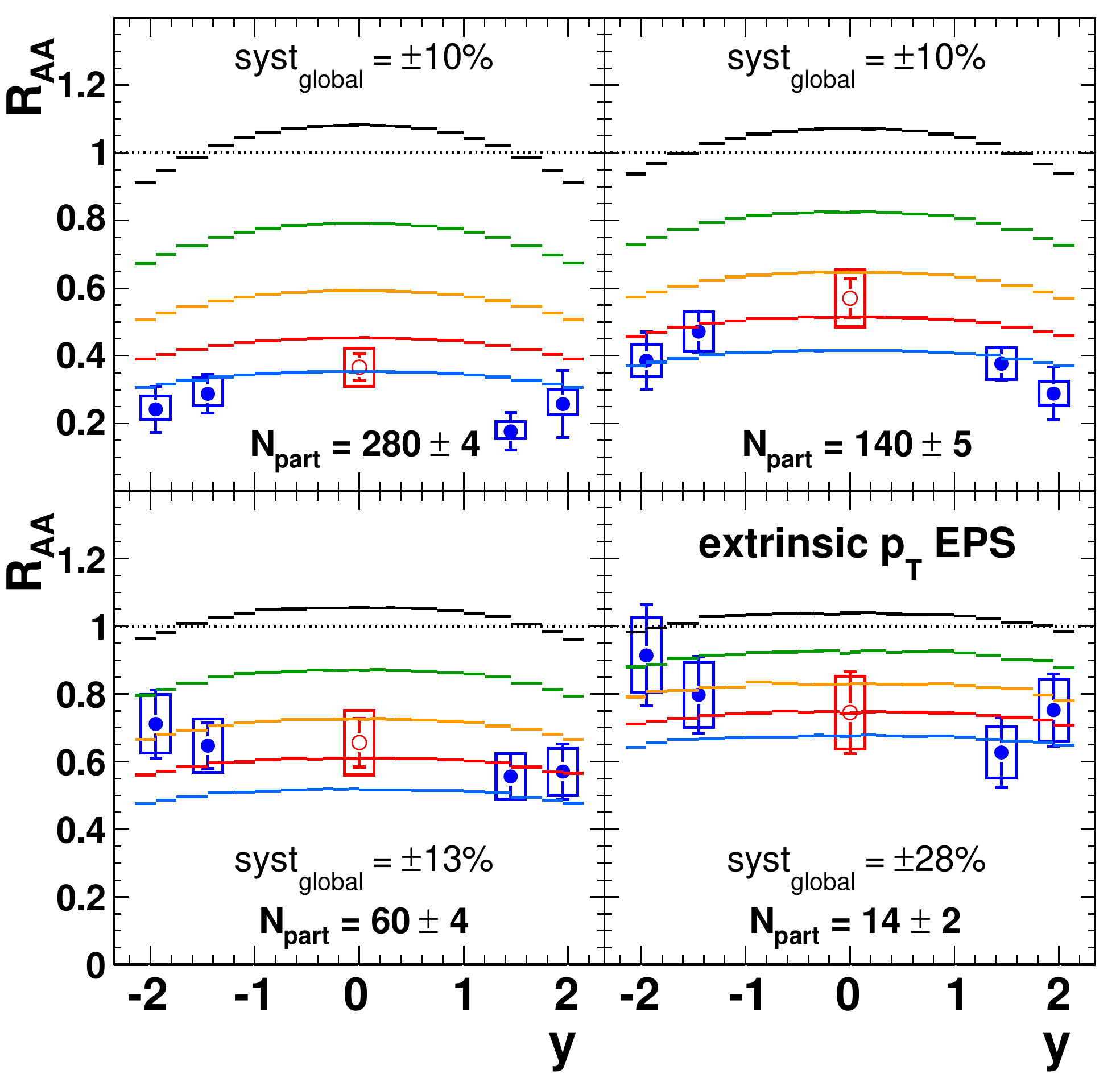}}
\subfigure[~nDSg]{\includegraphics[width=0.33\textwidth]{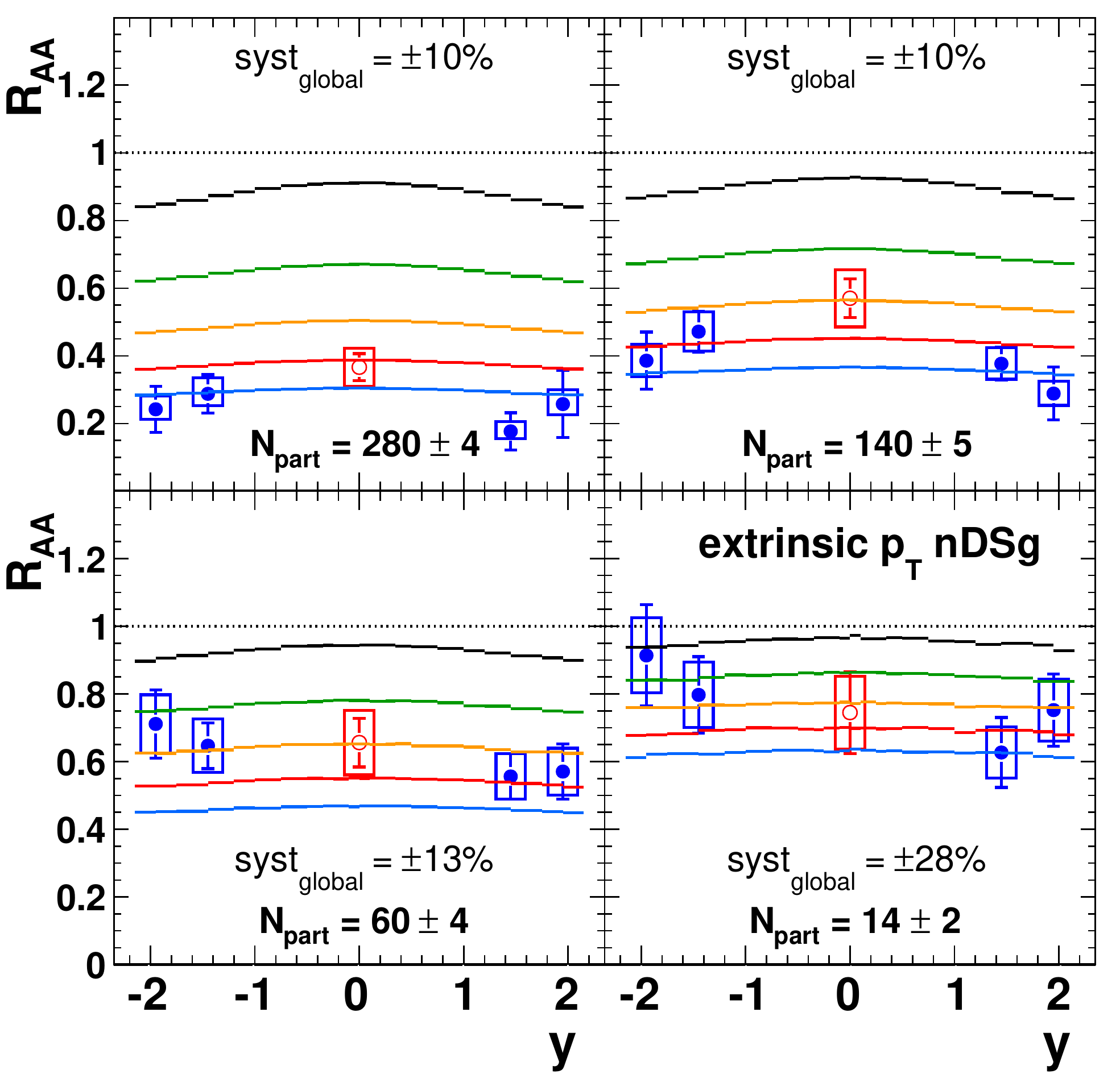}}
\caption{$R_{AA}$ vs $y$ for AuAu collisions for five $\sigma_{\rm abs}$ (from top to bottom: $\sigma_{abs}=$0, 2, 4, 6, 8 mb) using
 a) EKS98, b)  EPS08, c) nDSg in 4 centrality bins.}
\label{fig:R_AuAuvsy}
\subfigure[~EKS98]{\includegraphics[width=0.33\textwidth]{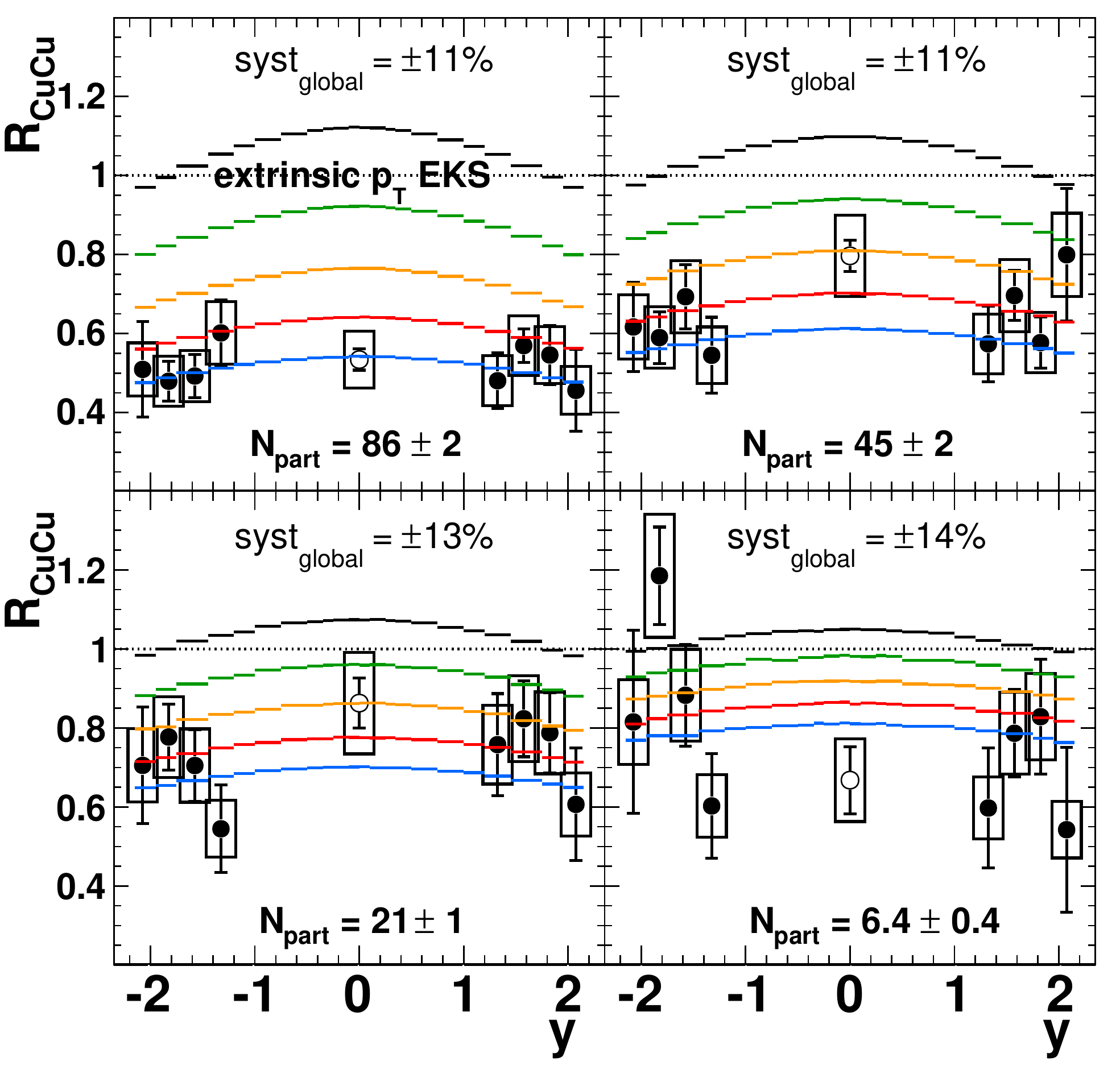}}
\subfigure[~EPS08]{\includegraphics[width=0.33\textwidth]{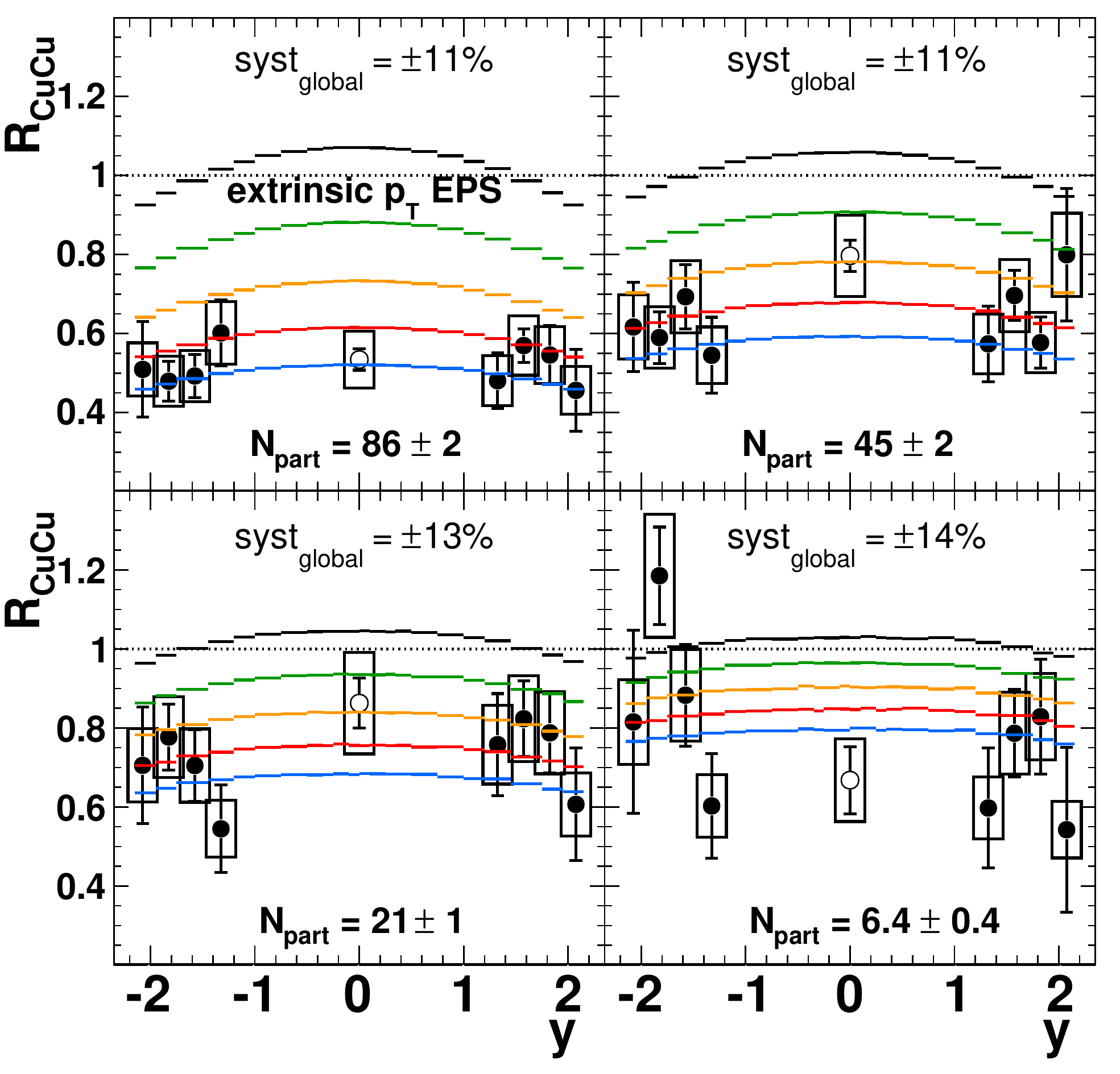}}
\subfigure[~nDSg]{\includegraphics[width=0.33\textwidth]{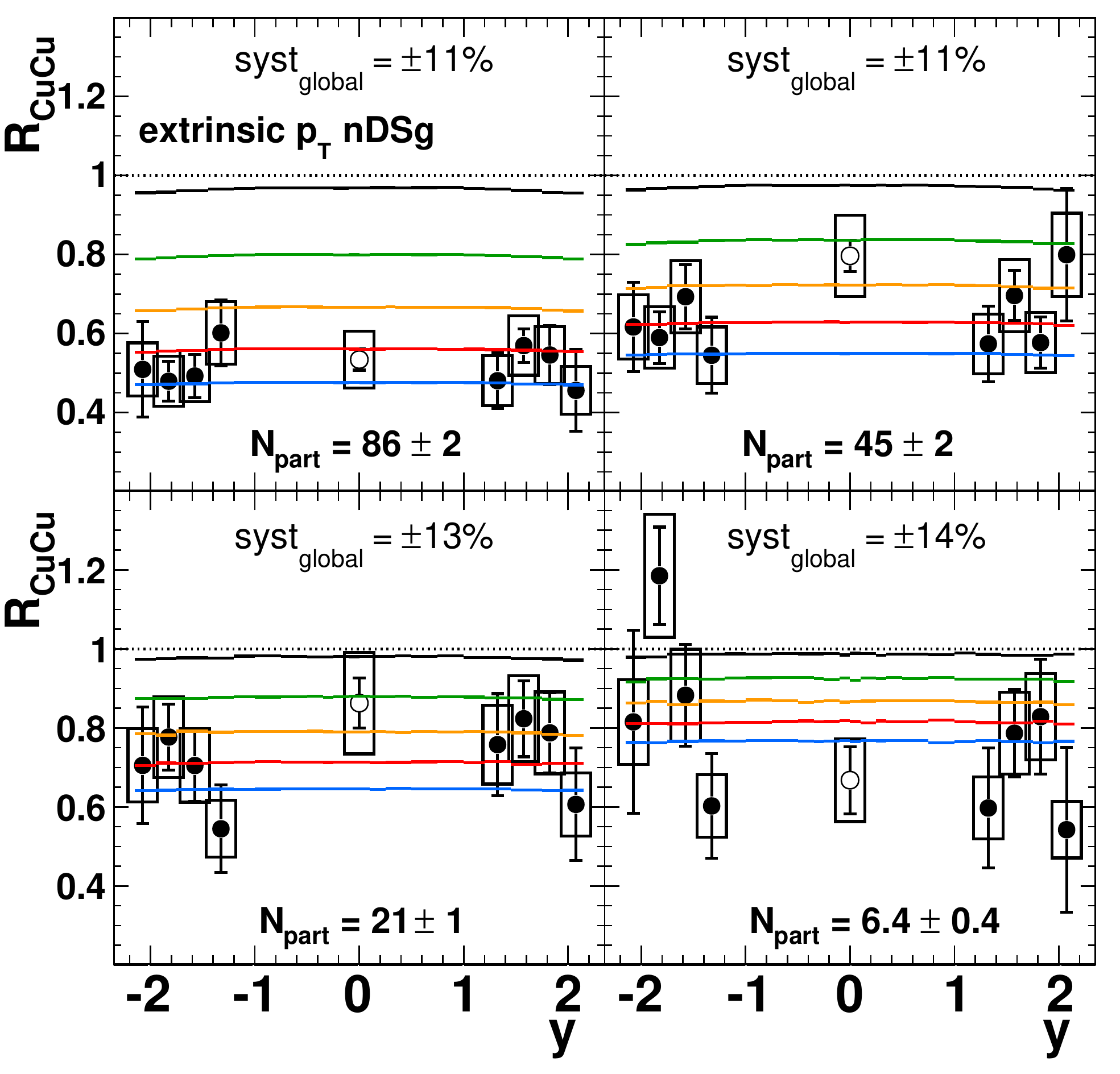}}
\caption{$R_{AA}$ vs $y$ for CuCu  collisions for five $\sigma_{\rm abs}$ (from top to bottom: $\sigma_{abs}=$0, 2, 4, 6, 8 mb) using
 a) EKS98, b)  EPS08, c) nDSg in 4 centrality bins.}
\label{fig:R_CuCuvsy}
\end{figure*}

\subsection{Rapidity dependence}

We now discuss the rapidity dependence of the nuclear modification factor in the case of CuCu and AuAu collisions.
It is displayed on~\cf{fig:R_AuAuvsy} and~\cf{fig:R_CuCuvsy} for the three shadowing parametrisations and for five $\sigma_{\rm abs}$.
As in~\cite{OurExtrinsicPaper}, $R_{AA}$ slightly peaks at $y=0$ reducing the need for recombination 
effects~\cite{recombinationRefs} which are particularly concentrated in the mid rapidity region and which
could elegantly explain  the differences of $R_{AA}$ between the forward and mid rapidity regions.

As we have noted in the previous section, shadowing effects exhibit naturally such a rapidity dependence. This happens for 
the three shadowing parametrisations we have used for \AuAu, confirming that this is a feature, rather than an accident. This effect 
is, however, reduced when an absorption cross section is accounted for. It is widely accepted that HNM effects are responsible for an 
extra suppression. If this suppression  -- by the creation of a QGP for instance -- is not correlated to the path of the $c \bar c$
on its way out of the nuclei, it may not damp down the difference between $R^{forward}_{AA}$ and $R^{central}_{AA}$, 
as does a larger $\sigma_{\rm abs}$.

\subsection{Transverse-momentum dependence}
\label{subsec:RAAvsPT}

 \begin{figure*}[thb!]
 \subfigure[~EKS98]{\includegraphics[width=0.33\textwidth]{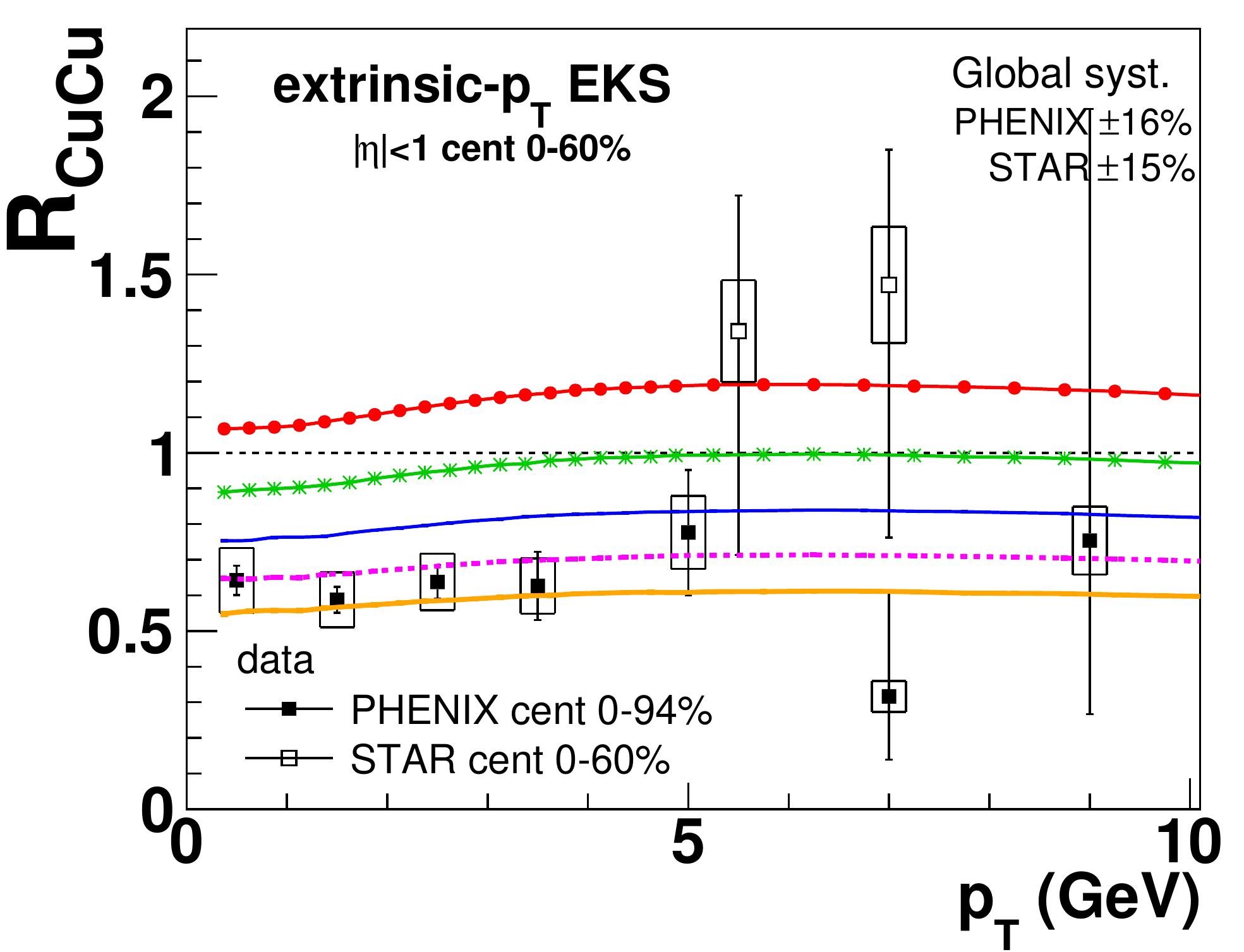}}
 \subfigure[~EPS08]{\includegraphics[width=0.33\textwidth]{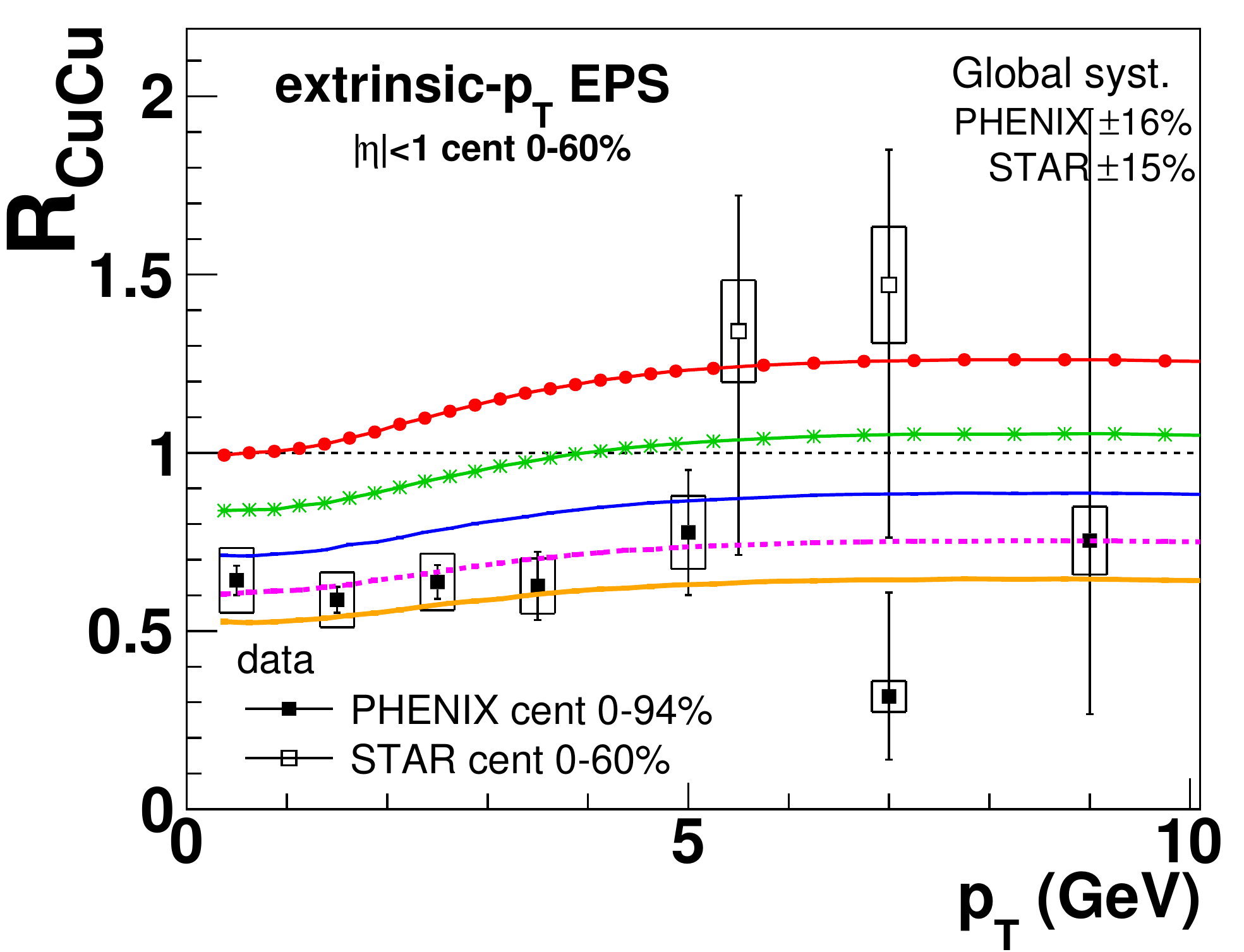}}
 \subfigure[~nDSg]{\includegraphics[width=0.33\textwidth]{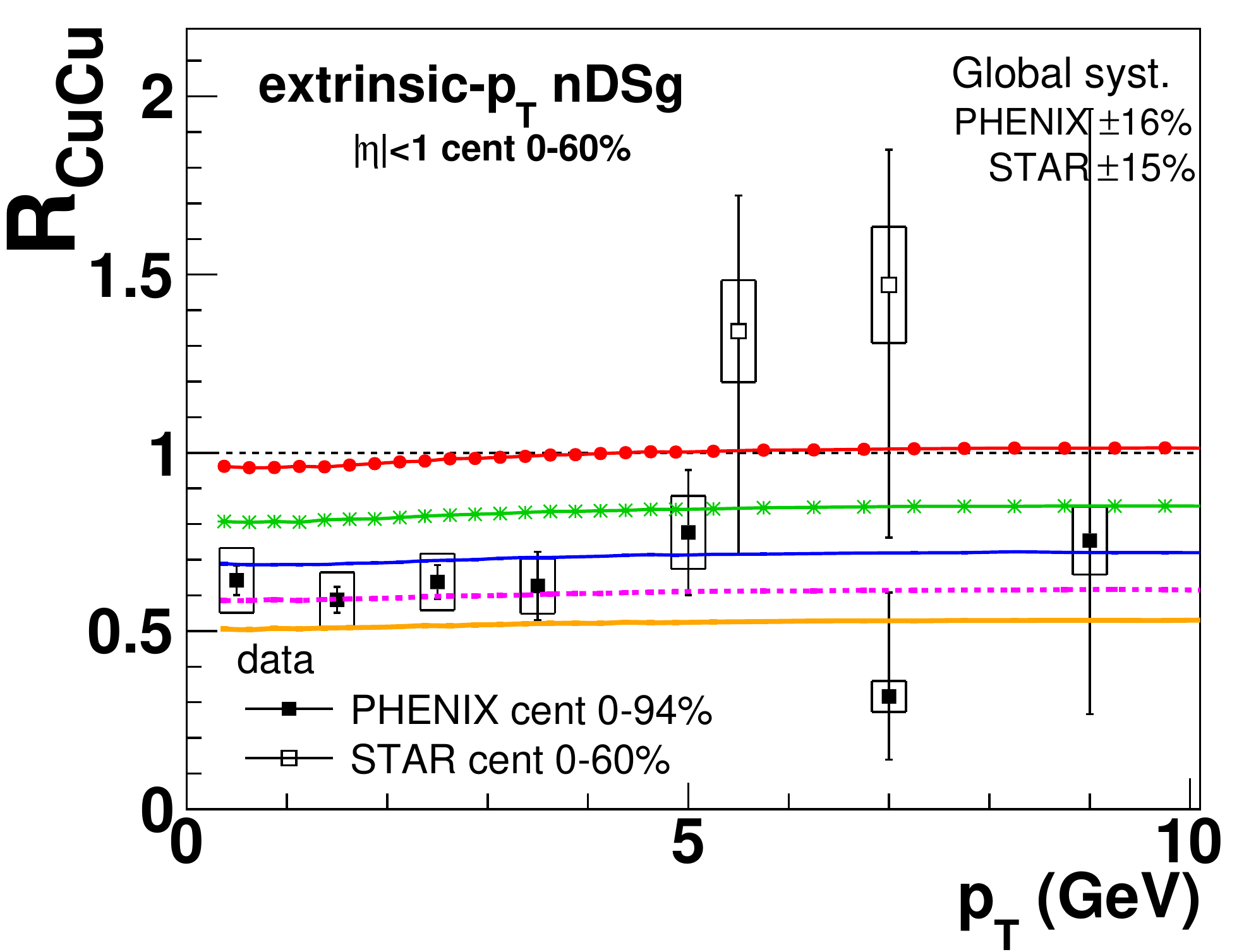}}
 \caption{(Color online) \jpsi\ nuclear modification factor in CuCu collisions at $\sqrt{s_{NN}}=200\mathrm{~GeV}$ versus $P_T$
 for the mid rapidity region for five values of the nuclear absorption (from top to bottom: $\sigma_{abs}=$0, 2, 4, 6, 8 mb)
  using : a) EKS98, b)  EPS08, c) nDSg. Those are compared with the two STAR points~\cite{Abelev:2009qaa} and
the PHENIX preliminary results~\cite{Leitch:2009xt}.}
 \label{fig:rCuCu_pT}
 \end{figure*}

We now move on to the discussion of the transverse momentum dependence in the mid rapidity region, 
analysed both by the PHENIX~\cite{Leitch:2009xt} and STAR~\cite{Abelev:2009qaa} collaborations. As announced during the
discussion of the dAu results, $R_{AA}$ versus $P_T$ increases with $P_T$ (see~\cf{fig:rCuCu_pT}).
In fact, the growth partially matches the trend of the PHENIX and STAR data. 
We should, however, mention here that there is no consensus
for now on whether one should expect a  nuclear modification factor
larger than one for $P_T$ around 8 GeV as seems to indicate the published STAR results~\cite{Abelev:2009qaa}. 

In the case of a confirmation of a non-suppression of $J/\psi$ at large $P_T$, one could say that it does not behave
as the other hadrons,  which are significantly suppressed in central Cu+Cu collisions and for increasing $P_T$ 
(see~\cite{adare:2008cx} for $\pi^0$ and~\cite{Garishvili:2009ei} for ``heavy-flavour'' muons)
In fact, 
the $J/\psi$ seems to adopt a behaviour
closer to the one of prompt photons~\cite{Reygers:2008pq} than to the one of other (heavy-flavoured) hadrons. 
We also note a non-suppression of $\phi$ meson in central Cu+Cu collisions~\cite{Naglis:2009uu}.

If one goes further, one may want to extract information about the production mechanism
at work here. Indeed, although the energy loss of a colored object in CNM is 
limited to be constant, rather than scaling with energy,  by the 
Landau-Pomeranchuk-Migdal effect~\cite{Brodsky:1992nq}, its magnitude per unit of length will
be significantly larger for a CO than for a CS state propagating in the nuclear matter.
This is especially relevant since, in the mid rapidity region, the $c \bar c$ pair hadronises outside the nucleus.
This would naturally lead us to the conclusion that it is rather a colourless state than a coloured one 
which propagates in the nuclear matter.

\section{Extraction of break-up cross section by fits on \dAu\ data}
\label{sec:fit}

By comparing our results in the intrinsic and extrinsic approaches, we have learnt that 
one of the consequences of this kinematical change implies a shift of the rapidity distribution. The latter is 
shifted as a whole to larger values of rapidity in the extrinsic case. As usual a $J/\psi$ break-up cross section, 
 $\sigma_{\mathrm{abs}}$, has to be accounted for to describe the normalisation of $R_{dAu}$. In practice, it 
is fit from the data and then used in the description of nucleus-nucleus collisions.
In view of the differences of the shadowing impact induced by one or the other kinematics, it is natural to wonder what
 the corresponding variations of $\sigma_{\mathrm{abs}}$ fit to the data are.

\subsection{Fitting $R_{dAu}$ data}

For this purpose, we have used PHENIX measurements of \RdAu~\cite{Adare:2007gn} in order to obtain the best
fit of $\sigma_{\mathrm{abs}}$ for each of the shadowing parametrisation considered
in both the intrinsic and extrinsic schemes. Based on  the method used by PHENIX 
in~\cite{Adare:2007gn} and \cite{Adare:2008cg}, we have computed  the $\chi^2$ in the different 
cases, including both statistical and systematic errors.

By using the data on  \RdAu\ versus rapidity, we have obtained the values of
$\sigma_{\mathrm{abs}}$ given 
in~\ct{tab:tab-1} for each of the shadowing parametrisations and for both production schemes.
The resulting curves are shown on \cf{fig:RdAu_vs_y} for dAu, on \cf{fig:R_AuAuvsybs} and \cf{fig:R_CuCuvsybs} for AuAu and CuCu.

\begin{table}[hbt!]
\begin{center}
\begin{tabular}{c|cc}
\hline
& ~~~~ $\sigma_{\mathrm{abs}}$ & $\chi^2_{min}$ \\ 
\hline\hline 
EKS98 Int. & $3.2 \pm 2.4$ & 0.9 \\
EPS08 Int. & $2.1^{+2.6}_{-2.2}$ & 1.1 \\
nDSg  Int. & $2.2^{+2.3}_{-2.1}$ & 1.3\\
\hline 
EKS98 Ext. & $3.9^{+2.7}_{-2.3}$ & 1.1 \\
EPS08 Ext. & $3.6^{+2.4}_{-2.5}$ & 0.5 \\
nDSg  Ext. & $3.0^{+2.2}_{-2.4}$ & 1.2 \\
\hline
\end{tabular}
\end{center}
\caption{$\sigma_{\mathrm{abs}}$ extracted from fit of $R_{dAu}$ (All cross section in units of mb)}
\label{tab:tab-1}
\end{table}

In general, fitting the rapidity with a constant $\sigma_{\mathrm{abs}}$ leads to more than acceptable $\chi^2$. The largest ones
are obtained for the nDSg parametrisation; this confirms the impression that parametrisation with ``significant'' shadowing and/or anti-shadowing
(EKS98, EPS08) are preferred by the data (see also \cf{fig:RdAu_vs_y}). The only systematic one can really see is that the 3  $\sigma_{\mathrm{abs}}$ 
values extracted using the extrinsic scheme are larger than the corresponding ones obtained with the intrinsic scheme, as expected due to the increase of $x_2$ in the extrinsic case compared to the intrinsic one. For the time being, 
the data are not precise enough to draw further conclusions from such fits.

\begin{figure*}[thb!]
\subfigure[~EKS98]{\includegraphics[width=0.33\textwidth]{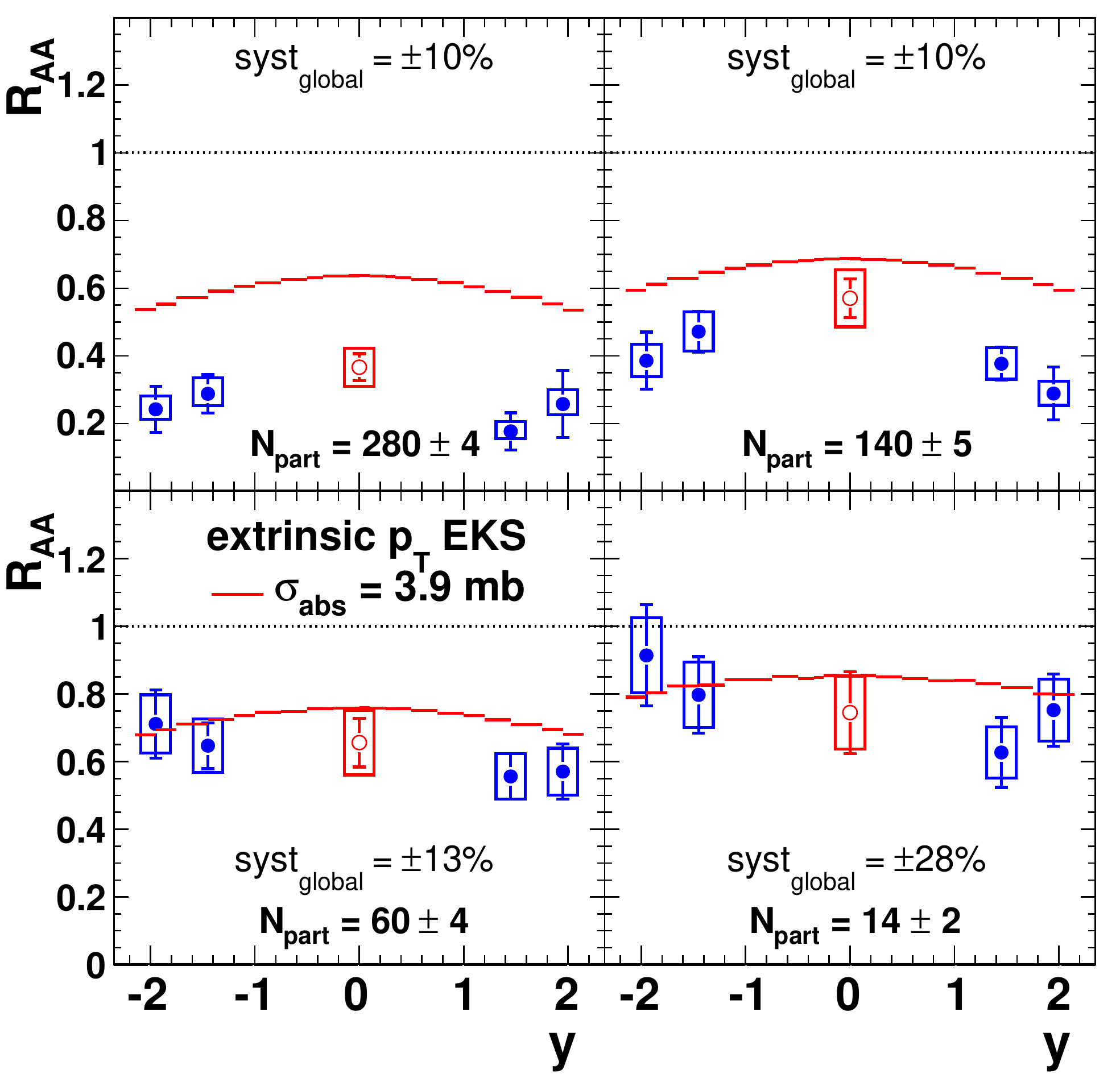}}
\subfigure[~EPS08]{\includegraphics[width=0.33\textwidth]{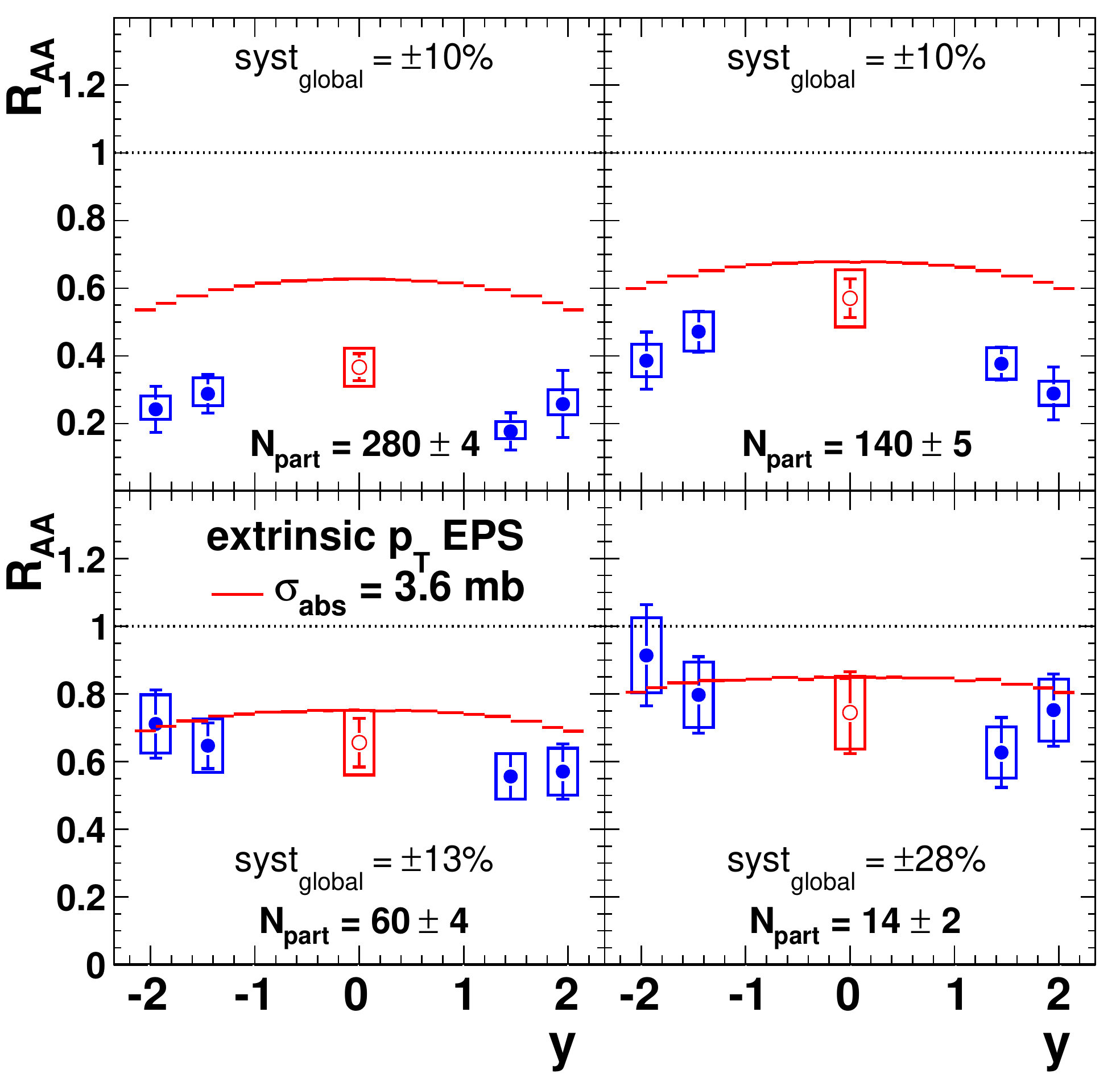}}
\subfigure[~nDSg]{\includegraphics[width=0.33\textwidth]{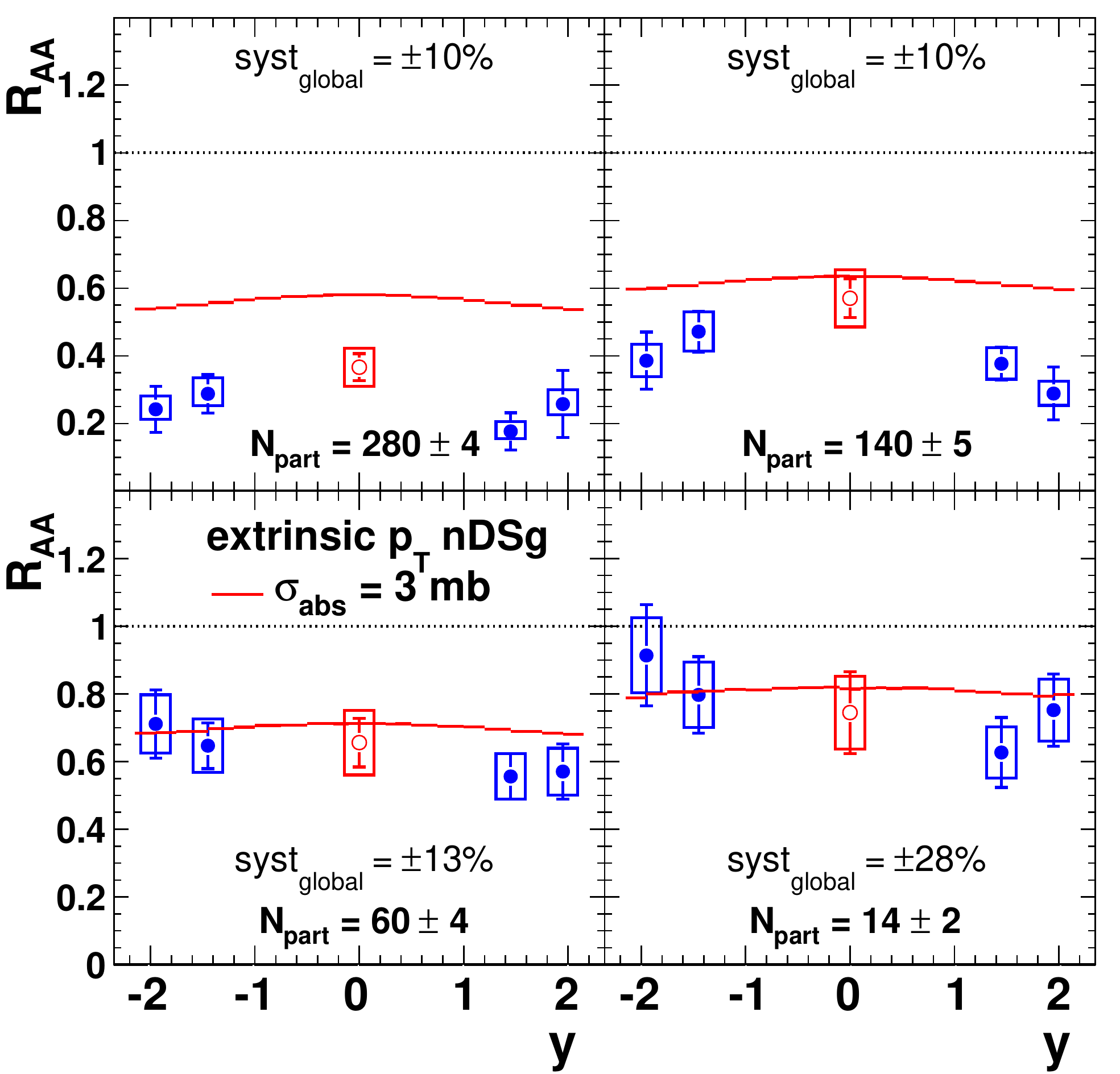}}
\caption{(Color online) $R_{AA}$ vs $y$ for AuAu collisions for the best $\sigma_{\rm abs}$ as obtained from the fit to the dAu data using a) EKS98, b)  EPS08, c) nDSg in 4 centrality bins.}
\label{fig:R_AuAuvsybs}
\subfigure[~EKS98]{\includegraphics[width=0.33\textwidth]{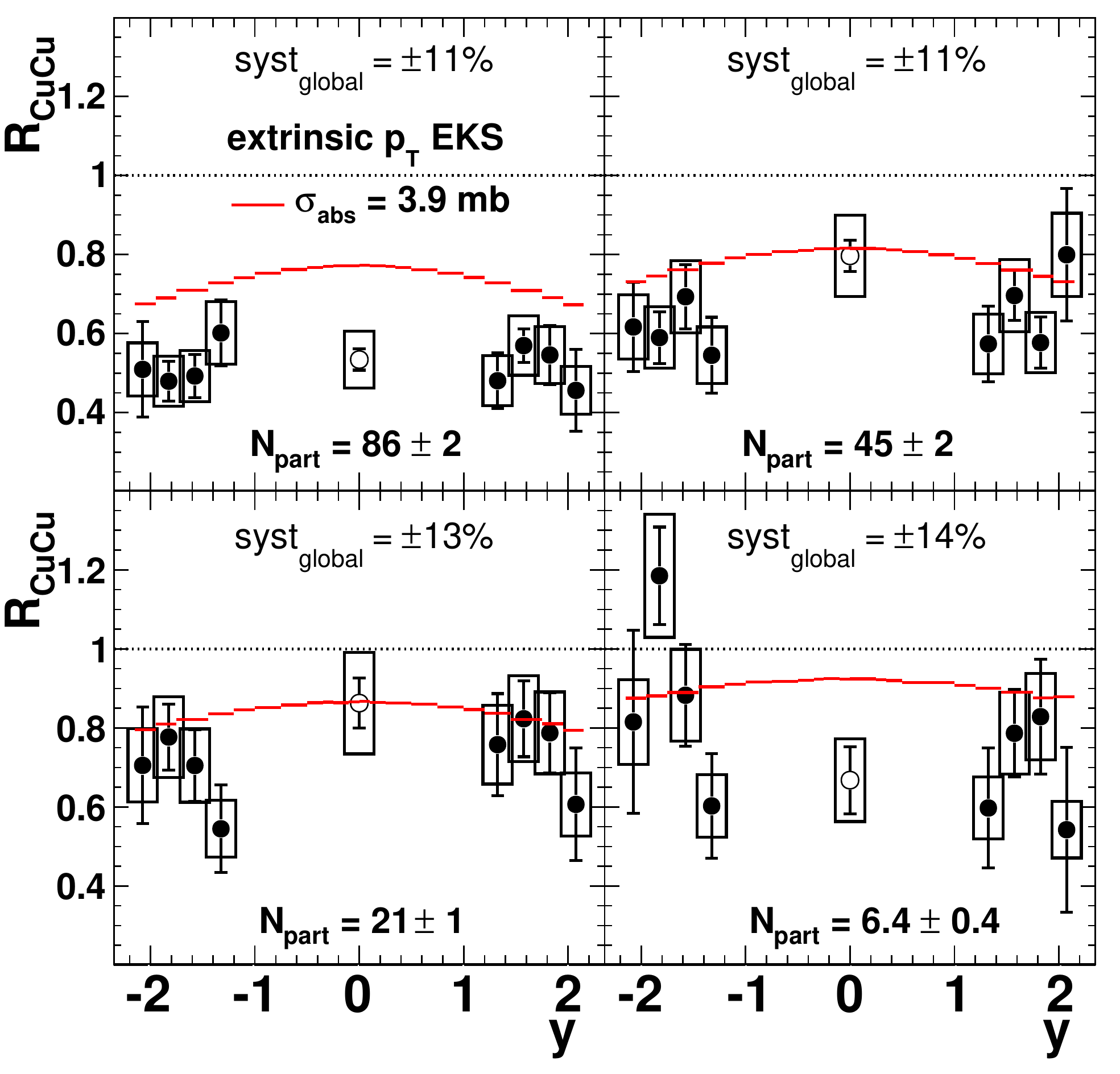}}
\subfigure[~EPS08]{\includegraphics[width=0.33\textwidth]{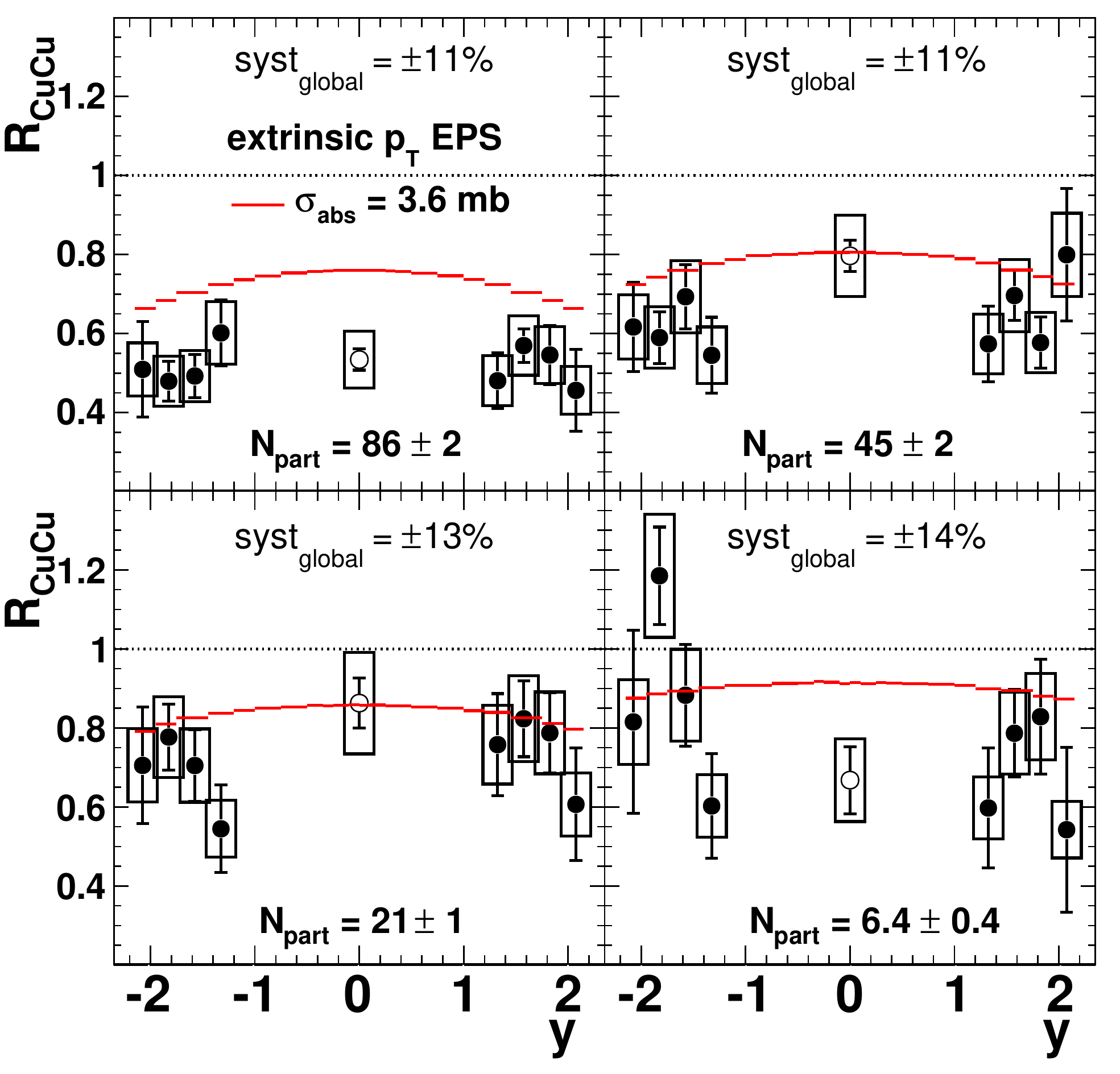}}
\subfigure[~nDSg]{\includegraphics[width=0.33\textwidth]{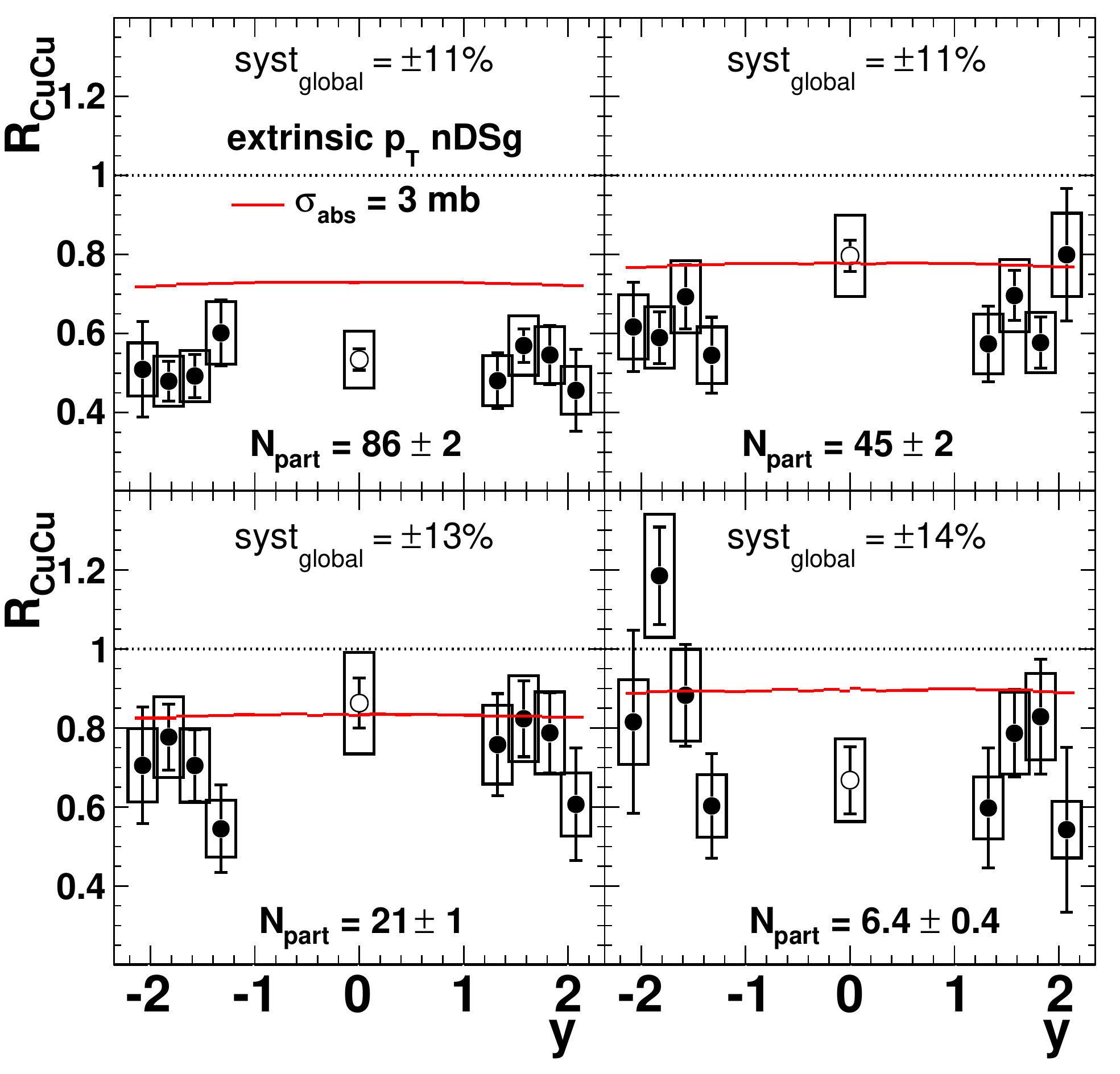}}
\caption{(Color online) $R_{AA}$ vs $y$ for CuCu  collisions for the best $\sigma_{\rm abs}$ as obtained from the fit to the dAu data using a) EKS98, b)  EPS08, c) nDSg in 4 centrality bins.}
\label{fig:R_CuCuvsybs}
\end{figure*}

\subsection{Fitting $R_{CP}$ data}

We have also fit the new $R_{CP}$ data with a constant $\sigma_{abs}$ in each
rapidity regions. Our results are given in~\ct{tab-2} along with the one 
of~\cite{frawley-INT} based on a $2\to 1$ kinematics and using the EKS98 
shadowing. We need to make clear at this point that the intrinsic scheme used 
in~\cite{frawley-INT} slightly differs from the one we have used for instance
to fit the dAu data as shown in the previous section. The difference appears at the level of 
the running of the scale of the shadowed gluon distribution and the invariant mass of 
produced system. However, this is not 
expected to modify the following discussion (see~\cite{OurIntrinsicPaper}).

\begin{table}[thb!]
\begin{tabular}{c|cccc}
\hline
& ~~~~ $y < 0$ ~~~~&  ~~~~ $y=0$ ~~~~& ~~~~ $y > 0$ ~~~~& ~~~~ all $y$~~~~~ \\ 
\hline\hline 
EKS98 Int.~\cite{frawley-INT} & $5.2 \pm 1.2$ & $3.1 \pm 1.3$  & $9.5 \pm 1.4$  & N/A \\
\hline 
EKS98 Ext. & $2.5 \pm 0.5$  & $3.2 \pm 0.5$  & $4.8 \pm 0.7$ & $3.2 \pm 0.4$ \\
EPS08 Ext. & $3.2 \pm 0.5$  & $2.5 \pm 0.5$  & $3.1 \pm 0.6$ & $2.9 \pm 0.3$ \\
nDSg  Ext. & $1.4 \pm 0.5$  & $1.6 \pm 0.5$  & $4.0 \pm 0.7$ & $2.2 \pm 0.3$ \\
\hline
\end{tabular}
\caption{$\sigma_{\mathrm{abs}}$ and their errors\protect\footnote{The errors quoted for the first line are extracted 
differently than ours. The errors shown here should only be compared for a given analysis.} extracted from fit 
of the $R_{CP}$ data averaged on three rapidity region as well as on the entire range. (All cross section in units of mb).}
\label{tab-2}
\end{table}

\begin{figure*}[thb!]
\begin{center}
\subfigure[~EKS98]{\includegraphics[width=.3\textwidth]{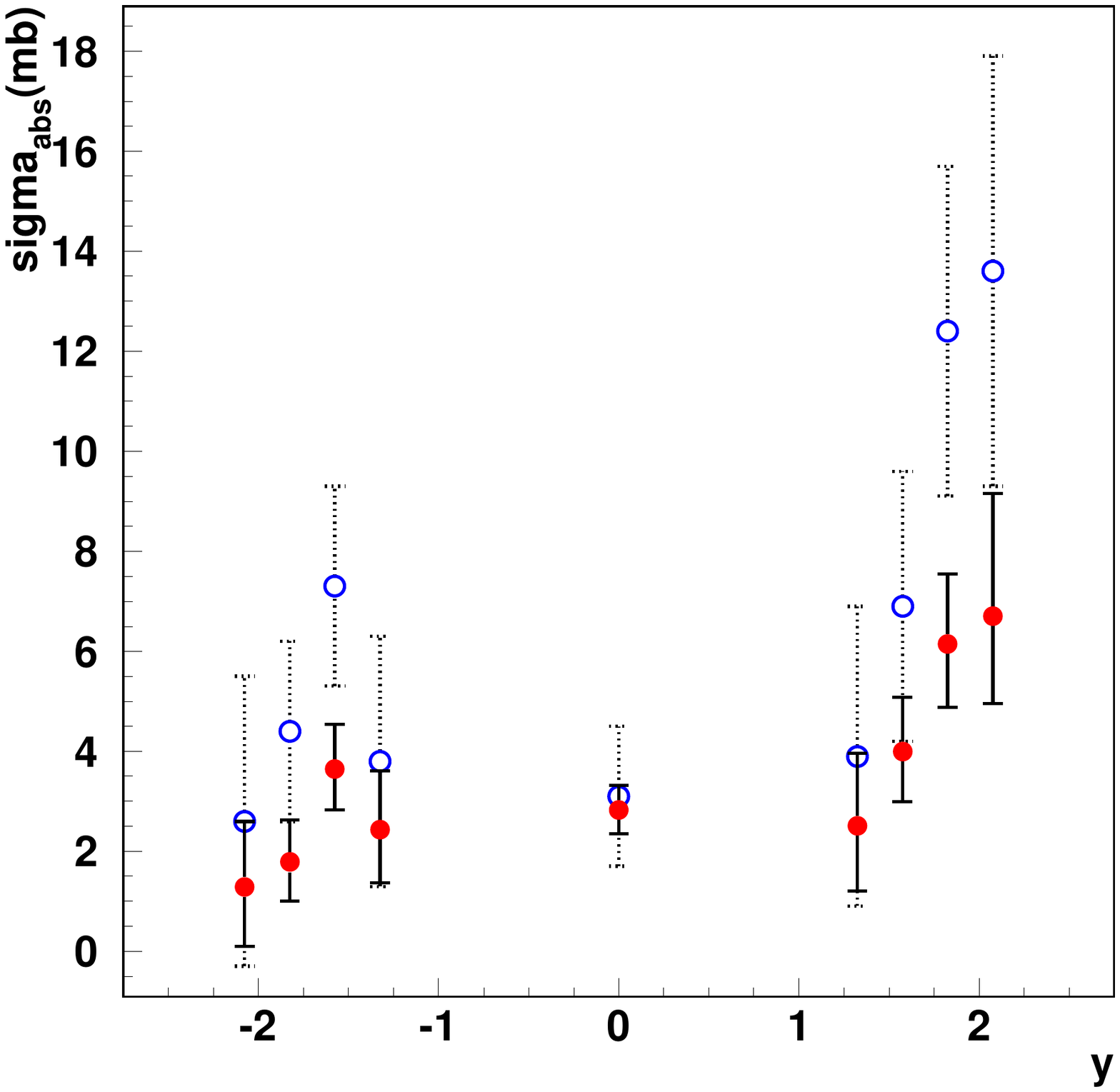}}
\subfigure[~EPS08]{\includegraphics[width=.3\textwidth]{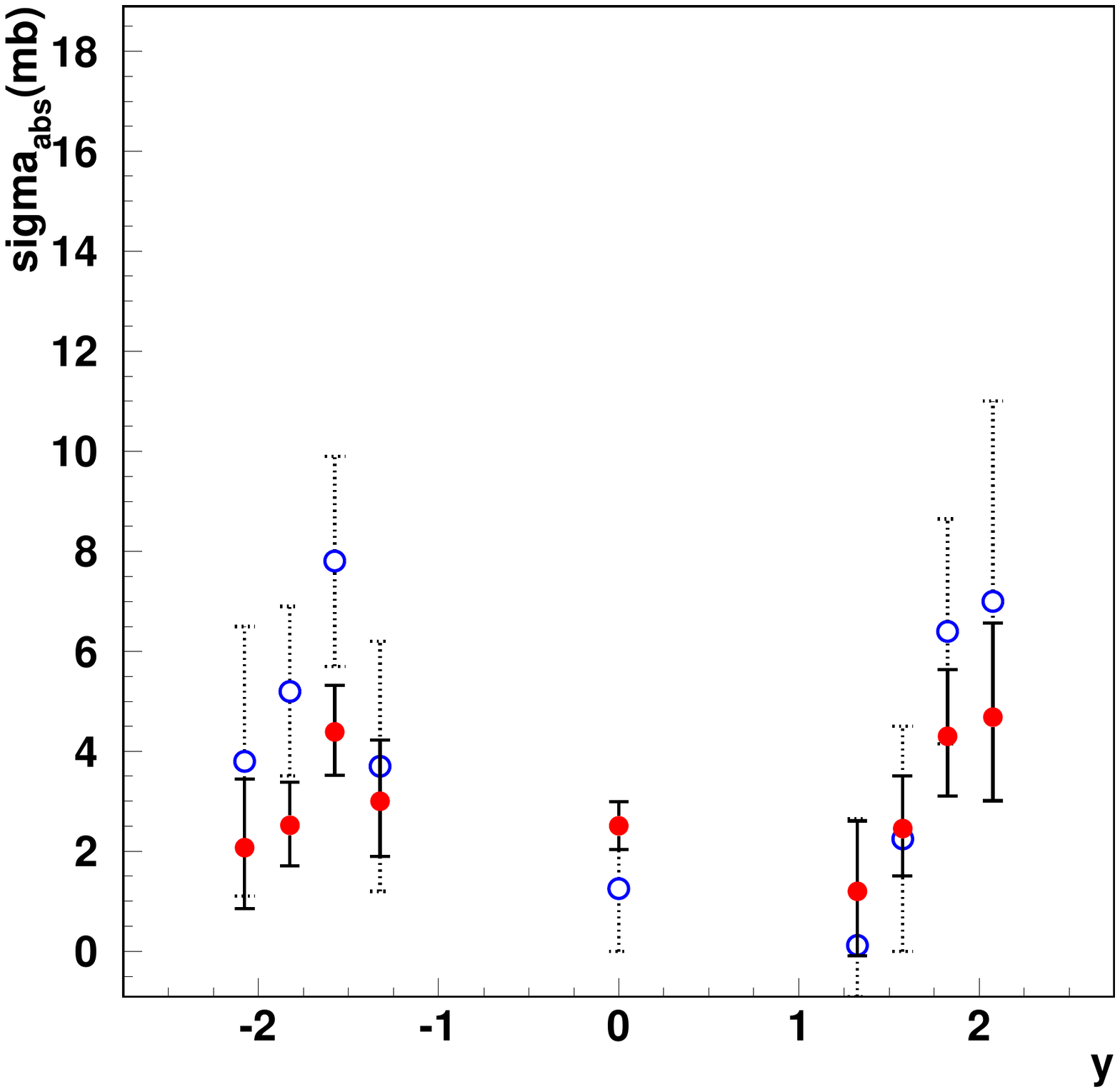}}
\subfigure[~nDSg]{\includegraphics[width=.3\textwidth]{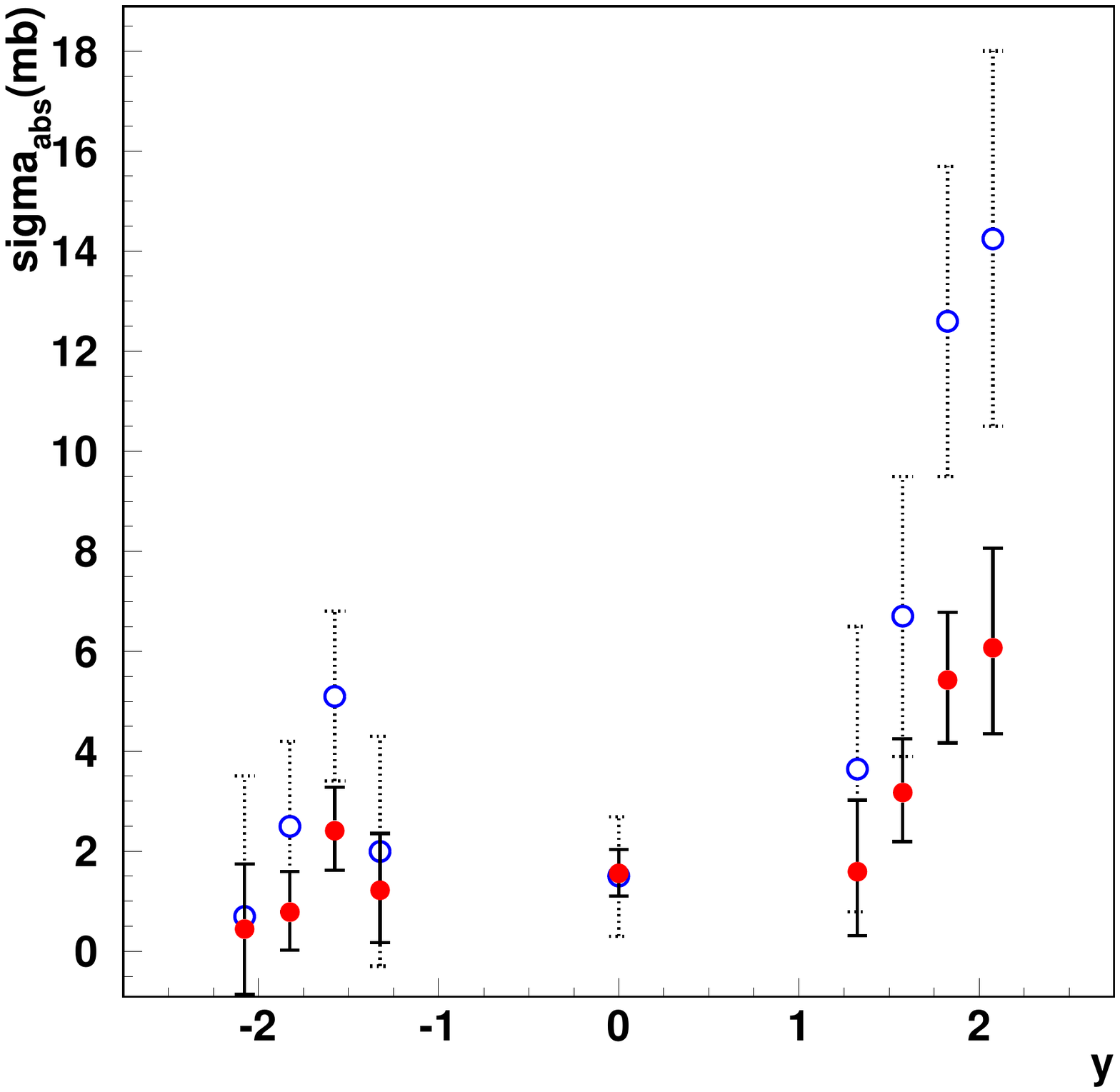}}
\end{center}
\caption{(Color online) $\sigma_{abs}$ versus $y$ obtained by fitting the $R_{CP}$ data for \dAu, in the extrinsic scheme (in red closed circles) compared to the intrinsic scheme~\cite{frawley-INT} (in blue open circles) using a) EKS98, b) EPS08  and c) nDSg.}
\label{fig:sigmaRCP}
\end{figure*}


As can be seen in~\ct{tab-2}, it appears
that the strong suppression at forward rapidity and the lack of suppression at
backward rapidity cannot be described using a fixed breakup cross section with EKS98 in the 
intrinsic scheme~\cite{frawley-INT}. One also seems to observe an increase 
of $\sigma_{abs}$ with rapidity in our analysis (extrinsic) for EKS98 and nDSg, but a constant behaviour 
cannot be ruled out. The increase is in any case softer when the final-state gluon momentum 
is taken into account. It seems that forward rapidity is maybe the most interesting region
for such investigations. Interestingly, in the case of EPS08, the value we have extracted for the 
forward region is equal to the one for the backward region. 

As we have argued earlier, EPS08 can be used as a good indicator of the strongest
shadowing reachable within the uncertainty of EPS09. From this viewpoint, 
the recent update of the intrinsic analysis, where~\cite{frawley-BNL} no
increase of $\sigma_{abs}$ is observed for the strongest shadowing of EPS09, confirms our findings.

To get a more precise view on the situation, it is useful
to plot the effective absorptive cross section as function of the rapidity, without averaging on the three rapidity regions.
The result can be seen on Figs. \ref{fig:sigmaRCP} with our result in the extrinsic scheme
(red closed circles) and the one of~\cite{frawley-INT} (blue open circles)\footnote{We recall here that the error bars on Figs. \ref{fig:sigmaRCP} are extracted with two different procedures in both case. They should only be compared within one approach.} in the intrinsic scheme.  We have obtained an increasing absorptive cross section with rapidity, but the increase is much less pronounced than in the intrinsic case. In fact, in the EPS08 case, the increase does not appear statistically significant.

\section{Conclusion}

Taking advantage of the probabilistic Glauber Monte-Carlo framework, {\sf JIN}, 
discussed in~\cite{OurIntrinsicPaper,OurExtrinsicPaper}, we have (i) 
considered three different gluon shadowing parametrisations -- EKS98, EPS08 and nDSg -- taking into
account a dependence on the impact parameter $b$ and the momentum of the gluon recoiling
against the $J/\psi$, (ii) shown
that the rapidity dependence of $R_{dAu}$ is shifted towards larger rapidities irrespective
of the shadowing parametrisation, and particularly the anti-shadowing peak, (iii) shown
that the anti-shadowing peak 
is reflected in a rise of the nuclear modification factor for
increasing $P_T$, (iv)  compared our results with the experimental
measurements of the nuclear modification factors $R_{dAu}$ and
$R_{CP}$ from \dAu\ collisions presently available at RHIC and 
extracted the favoured values of the $c\bar c$ absorption cross section
in the nuclear matter, and finally (v) shown that the effective absorption 
cross section increase at forward rapidity, obtained from the recent analysis of 
PHENIX $R_{CP}$ data~\cite{frawley-INT} in which the final-state gluon momentum is neglected (intrinsic case), 
is less marked when it is taken into account; that is in the extrinsic case.


\section*{Acknowledgments}


We would like to thank S.J. Brodsky, A. Linden-Levy, C. Louren\c co, N. Matagne, J. Nagle, T.~Ullrich, R.~Vogt, H.~W\"ohri for stimulating and useful discussions. This work is supported in part by Xunta de Galicia (2008/012) and Ministerio de Educacion y Ciencia of Spain (FPA2008-03961-E/IN2P3), the Belgian American Educational Foundation, the Francqui Foundation and the U.S. Department of Energy under contract number DE-AC02-76SF00515.



\begin{thebibliography}{99}

\bibitem{Matsui86}
  T.~Matsui and H.~Satz,
  Phys.\ Lett.\  B {\bf 178} (1986) 416.

\bibitem{Adare:2006ns}
  A.~Adare {\it et al.},
  Phys.\ Rev.\ Lett.\  {\bf 98}, 232301 (2007).

\bibitem{Adare:2007gn}
  A.~Adare {\it et al.}, 
  Phys.\ Rev.\  C {\bf 77} (2008) 024912.

\bibitem{Frawley:2008kk}
  A.~D.~Frawley, T.~Ullrich and R.~Vogt,
  Phys.\ Rept.\  {\bf 462} (2008) 125;

\bibitem{Rapp:2008tf}
  R.~Rapp, D.~Blaschke and P.~Crochet,
   arXiv:0807.2470;

\bibitem{Kluberg:2009wc}
  L.~Kluberg and H.~Satz,
   arXiv:0901.3831.


\bibitem{Lansberg:2008zm}
  J.~P.~Lansberg {\it et al.},
  AIP Conf.\ Proc.\  {\bf 1038} (2008) 15.

\bibitem{OurIntrinsicPaper}
  E.~G.~Ferreiro, F.~Fleuret and A.~Rakotozafindrabe,
  Eur.\ Phys.\ J.\  C {\bf 61} (2009) 859.


\bibitem{OurExtrinsicPaper}
  E.~G.~Ferreiro, F.~Fleuret, J.~P.~Lansberg and A.~Rakotozafindrabe,
  Phys.\ Lett.\  B {\bf 680}, 50 (2009).

\bibitem{Perkins:2009tp}
  C.~Perkins  [STAR Collaboration],
  Nucl.\ Phys.\  A {\bf 830} (2009) 231C.

\bibitem{daSilva:2009yy}
  C.~L.~da Silva  [PHENIX Collaboration],
  Nucl.\ Phys.\  A {\bf 830} (2009) 227C.

\bibitem{Hodgson:1971}
P.~E. Hodgson, {\it Nuclear Reactions and Nuclear Structure}, Clarendon Press, (1971) 453 p

\bibitem{Lansberg:2006dh}
  J.~P.~Lansberg, Int.\ J.\ Mod.\ Phys.\ A {\bf 21} (2006) 3857

\bibitem{Haberzettl:2007kj}
  H.~Haberzettl and J.~P.~Lansberg,
  Phys.\ Rev.\ Lett.\  {\bf 100} (2008) 032006.

\bibitem{Brodsky:2009cf}
  S.~J.~Brodsky and J.~P.~Lansberg,
  Phys.\ Rev.\  D {\bf 81} 051502 (R) (2010).

\bibitem{Adare:2006kf}
  A.~Adare {\it et al.}, 
  Phys.\ Rev.\ Lett.\  {\bf 98} 232002 (2007) .

\bibitem{CSM_hadron}
C-H. Chang,
{Nucl. Phys. } B {\bf 172} (1980) 425;
  E.~L.~Berger and D.~L.~Jones,
  Phys.\ Rev.\  D {\bf 23} (1981) 1521;
R. Baier and R. R\"uckl,
{Phys. Lett. } B {\bf 102} (1981) 364;
R. Baier and R. R\"uckl,
 {Z. Phys. } C {\bf 19} (1983) 251;
  V.~G.~Kartvelishvili, A.~K.~Likhoded and S.~R.~Slabospitsky,
  Sov.\ J.\ Nucl.\ Phys.\  {\bf 28} (1978) 678
  [Yad.\ Fiz.\  {\bf 28} (1978) 1315].

\bibitem{Campbell:2007ws}
  J.~M.~Campbell, F.~Maltoni and F.~Tramontano,
  Phys.\ Rev.\ Lett.\  {\bf 98} (2007) 252002.

\bibitem{Artoisenet:2007xi}
  P.~Artoisenet, J.~P.~Lansberg and F.~Maltoni,
  Phys.\ Lett.\  B {\bf 653} (2007) 60.

\bibitem{Gong:2008sn}
  B.~Gong and J.~X.~Wang,
  Phys.\ Rev.\ Lett.\  {\bf 100} (2008) 232001.

\bibitem{Artoisenet:2008fc}
  P.~Artoisenet, J.~M.~Campbell, J.~P.~Lansberg, F.~Maltoni and F.~Tramontano,
  Phys.\ Rev.\ Lett.\  {\bf 101} (2008) 152001.

\bibitem{Lansberg:2008gk}
  J.~P.~Lansberg,
  Eur.\ Phys.\ J.\  C {\bf 61} (2009) 693.

\bibitem{Gong:2009kp}
  B.~Gong and J.~X.~Wang,
  Phys.\ Rev.\ Lett.\  {\bf 102} (2009) 162003.



\bibitem{Ma:2008gq}
  Y.~Q.~Ma, Y.~J.~Zhang and K.~T.~Chao,
  Phys.\ Rev.\ Lett.\  {\bf 102} (2009) 162002.

\bibitem{Pakhlov:2009nj}
  P.~Pakhlov {\it et al.}  [Belle Collaboration],
  Phys.\ Rev.\  D {\bf 79} (2009) 071101.

\bibitem{Zhang:2009ym}
  Y.~J.~Zhang, Y.~Q.~Ma, K.~Wang and K.~T.~Chao,
  Phys.\ Rev.\  D {\bf 81} (2010) 034015;
  Z.~G.~He, Y.~Fan and K.~T.~Chao,
  Phys.\ Rev.\  D {\bf 81} (2010) 054036.



\bibitem{Cho:1995ce}
  P.~L.~Cho and A.~K.~Leibovich,
  Phys.\ Rev.\  D {\bf 53} (1996) 6203.

\bibitem{frawley-INT} 
A. D. Frawley, {\it Update on $J/\psi$ cold nuclear matter $R_{AA}$
estimates from fits to \dAu\ $R_{CP}$ data}, talk at  Joint CATHIE-INT 
mini-program Quarkonia in Hot QCD, INT, Seattle, USA, June 16-26, 2009.

\bibitem{frawley-BNL} 
A. D. Frawley, {\it Cold nuclear matter effects on $J/\psi$ production}, talk at  Joint CATHIE/TECHQM Workshop
, BNL, Brookhaven, USA, December 14-18, 2009.

\bibitem{OtherShadowingRefs}
  T.~Gousset and H.~J.~Pirner,
  Phys.\ Lett.\  B {\bf 375} (1996) 349; 
  %
  R.~Vogt,
  Phys.\ Rev.\  C {\bf 61} (2000) 035203; 
  %
  F.~Arleo and V.~N.~Tram,
  Eur.\ Phys.\ J.\  C {\bf 55} (2008) 449; 
  %
  F.~Arleo,
  Phys.\ Lett.\ B {\bf 666} (2008) 31; 
  %
  C.~Lourenco, R.~Vogt and H.~K.~Woehri,
  JHEP {\bf 0902} (2009) 014


\bibitem{Adare:2006kf}
  A.~Adare {\it et al.}, 
  Phys.\ Rev.\ Lett.\  {\bf 98} (2007) 232002.

\bibitem{Lansberg:2005pc}
  J.~P.~Lansberg, J.~R.~Cudell and Yu.~L.~Kalinovsky,
  Phys.\ Lett.\ B {\bf 633} (2006) 301.

\bibitem{Eskola:1998df}
  K.~J.~Eskola, V.~J.~Kolhinen and C.~A.~Salgado,
  Eur.\ Phys.\ J.\  C {\bf 9} (1999) 61.

\bibitem{Eskola:2008ca}
  K.~J.~Eskola, H.~Paukkunen and C.~A.~Salgado,
  JHEP {\bf 0807} (2008) 102.

\bibitem{deFlorian:2003qf}
  D.~de Florian and R.~Sassot,
  Phys.\ Rev.\  D {\bf 69} (2004) 074028.

\bibitem{Eskola:2009uj}
  K.~J.~Eskola, H.~Paukkunen and C.~A.~Salgado,
  JHEP {\bf 0904} (2009) 065.

\bibitem{Kopeliovich:2008}
  B.~Z.~Kopeliovich, E.~Levin, I.~K.~Potashnikova and I. Schmidt,
  Phys.\ Rev.\ C {\bf 79} (2009) 064906.

\bibitem{Kopeliovich:2005}
  B.~Z.~Kopeliovich, J.~Nemchik, I.~K.~Potashnikova, M.~B.~Johnson and I. Schmidt,
  Phys.\ Rev.\ C {\bf 72} (2005) 054606.
  

\bibitem{Klein:2003dj}
  S.~R.~Klein and R.~Vogt,
  Phys.\ Rev.\ Lett.\  {\bf 91} (2003) 142301.

\bibitem{Vogt:2004dh}
  R.~Vogt,
  Phys.\ Rev.\  C {\bf 71} (2005) 054902.

\bibitem{Adare:2008cg}
  A.~Adare {\it et al.}, 
  Phys.\ Rev.\  C {\bf 77} (2008) 064907.

\bibitem{Bedjidian:2004gd}
  M.~Bedjidian {\it et al.}, CERN-2004-009-C.

\bibitem{recombinationRefs}
  L.~Grandchamp, R.~Rapp and G.~E.~Brown,
  Phys.\ Rev.\ Lett.\  {\bf 92} (2004) 212301; 
  %
  E.~L.~Bratkovskaya, A.~P.~Kostyuk, W.~Cassing and H.~Stoecker,
  Phys.\ Rev.\  C {\bf 69} (2004) 054903; 
  %
  R.~L.~Thews,
  Eur.\ Phys.\ J.\  C {\bf 43} (2005) 97; 
  %
  L.~Yan, P.~Zhuang and N.~Xu,
  Phys.\ Rev.\ Lett.\  {\bf 97} (2006) 232301; 
  %
  A.~Andronic, P.~Braun-Munzinger, K.~Redlich and J.~Stachel,
  Nucl.\ Phys.\  A {\bf 789} (2007) 334; 
  A.~Capella, L.~Bravina, E.~G.~Ferreiro, A.~B.~Kaidalov, K.~Tywoniuk 
  and E.~Zabrodin,
  Eur.\ Phys.\ J.\  C {\bf 58}, 437 (2008).

\bibitem{Abelev:2009qaa}
  B.~I.~Abelev {\it et al.}  [STAR Collaboration],
  Phys.\ Rev.\  C {\bf 80} (2009) 041902.

\bibitem{Leitch:2009xt}
  M.~Leitch  [PHENIX Collaboration],
  Nucl.\ Phys.\  A {\bf 830} (2009) 27C.

\bibitem{adare:2008cx}
  A.~Adare {\it et al.}  [PHENIX Collaboration],
  Phys.\ Rev.\ Lett.\  {\bf 101} (2008) 162301.

\bibitem{Garishvili:2009ei}
  I.~Garishvili  [PHENIX Collaboration],
  Nucl.\ Phys.\  A {\bf 830} (2009) 625C.

\bibitem{Naglis:2009uu}
  M.~Naglis  [PHENIX Collaboration],
  Nucl.\ Phys.\  A {\bf 830} (2009) 757C.

\bibitem{Reygers:2008pq}
  K.~Reygers  [PHENIX Collaboration],
  J.\ Phys.\ G {\bf 35} (2008) 104045.

\bibitem{Adare:2008sh}
  A.~Adare {\it et al.},  
  Phys.\ Rev.\ Lett.\  {\bf 101} (2008) 122301
; S. Oda {\it et al.}, J.\ Phys.\ G {\bf 35}, 104134 (2008).

\bibitem{Brodsky:1992nq}
  S.~J.~Brodsky and P.~Hoyer,
  Phys.\ Lett.\  B {\bf 298} (1993) 165.

\expandafter\ifx\csname natexlab\endcsname\relax\def\natexlab#1{#1}\fi
\expandafter\ifx\csname bibnamefont\endcsname\relax
  \def\bibnamefont#1{#1}\fi
\expandafter\ifx\csname bibfnamefont\endcsname\relax
  \def\bibfnamefont#1{#1}\fi
\expandafter\ifx\csname citenamefont\endcsname\relax
  \def\citenamefont#1{#1}\fi
\expandafter\ifx\csname url\endcsname\relax
  \def\url#1{\texttt{#1}}\fi
\expandafter\ifx\csname urlprefix\endcsname\relax\def\urlprefix{URL }\fi
\providecommand{\bibinfo}[2]{#2}
\providecommand{\eprint}[2][]{\url{#2}}

\end{thebibliography}
\end{document}